\documentclass[manuscript]{aastex}

\usepackage{graphicx,epsfig,subfigure,rotating}
\usepackage{amssymb,amsmath}

\begin{document}


\title{Source Detection in Interferometric Visibility Data\\I. Fundamental Estimation Limits}


\author{Cathryn M. Trott}
\affil{Department of Radiology, Massachusetts General Hospital, Boston MA 02114, and\\
International Centre for Radio Astronomy Research, Curtin University, Bentley WA, Australia}
\email{ctrott@pet.mgh.harvard.edu}

\author{Randall B. Wayth}
\affil{International Centre for Radio Astronomy Research, Curtin University, Bentley WA, Australia}

\author{Jean-Pierre R. Macquart}
\affil{International Centre for Radio Astronomy Research, Curtin University, Bentley WA, Australia}

\author{Steven J. Tingay}
\affil{International Centre for Radio Astronomy Research, Curtin University, Bentley WA, Australia}




\begin{abstract}
Transient radio signals of astrophysical origin present an avenue for studying the dynamic universe. With the next generation of radio interferometers being planned and built, there is great potential for detecting and studying large samples of radio transients. Currently-used image-based techniques for detecting radio sources have not been demonstrated to be optimal, and there is a need for development of more sophisticated algorithms, and methodology for comparing different detection techniques. A visibility-space detector benefits from our good understanding of visibility-space noise properties, and does not suffer from the image artifacts and need for deconvolution in image-space detectors. In this paper, we propose a method for designing optimal source detectors using visibility data, building on statistical decision theory. The approach is substantially different to conventional radio astronomy source detection. Optimal detection requires an accurate model for the data, and we present a realistic model for the likelihood function of radio interferometric data, including the effects of calibration, signal confusion and atmospheric phase fluctuations. As part of this process, we derive fundamental limits on the calibration of an interferometric array, including the case where many relatively weak ``in-beam" calibrators are used. These limits are then applied, along with a model for atmospheric phase fluctuations, to determine the limits on measuring source position, flux density and spectral index, in the general case. We then present an optimal visibility-space detector using realistic models for an interferometer.
\end{abstract}




\section{Introduction}
The next generation of wide-field survey radio interferometers (MWA, ASKAP, ATA, LOFAR, LWA), culminating in the Square Kilometre Array (SKA), faces new challenges in meeting the ambitious science goals of the twenty-first century. These goals demand advances in telescope engineering and data-processing design, as well as sophisticated and novel observational techniques. Wide-field instruments will generate data at high rates, and therefore require techniques that optimally use the data. One of the main science goals of these instruments is to detect transient and variable radio sources, and as such a key requirement will be optimal source detection.

The time domain, historically dominated by pulsar observations, is broadening to study new classes of dynamic sources. New high-sensitivity, large collecting area instruments afford the opportunity to detect and study transient signals. One group of well-known radio transients are pulsars. The periodicity of these sources is used to detect them. The more general class of transients sources, with non-periodic behaviour (episodic), has not been surveyed and studied systematically. As well as the knowledge we can gain from studying expected transient sources (e.g., GRBs, pulsating stars), there is great potential for observing exotic astrophysical events such as annihilating black holes, gravity wave events (e.g., colliding black holes), magnetars and extraterrestrial signals \citep{jp10a}. Optimal source detection has a role for detection of both fast and slow transients.

Detection of signals in radio astronomy typically occurs in the image domain, and uses either a simple highest-peak thresholding, or a matched filtering operation \citep{cordes09}. The former sets a threshold value above the noise level, and attempts to detect the strongest signal in an image, and then model the incompleteness to remove its sidelobes. The next strongest signal is then compared to the threshold and modelled, and the process continues until there are no more signals above the threshold. This process works well for strong signals that are well-separated (no source confusion), and well-understood noise.

Matched filtering is an operation that is derived from statistical decision theory. The matched filter correlates the received data with a replica of the signal \citep{kay98}. It weights the data according to the signal strength, thereby allowing the datapoints with the strongest signals to contribute more to the filter output. The matched filter is optimal for white gaussian noise (uncorrelated noise) and known signal. For non-Gaussian noise, the matched filter is no longer optimal \citep{kay98}. Hence, the matched filter is not necessarily the best detector that can be designed for source detection in image-space data. In addition, to perform a matched filtering operation with sources with unknown parameters (e.g., sky position, strength), radio astronomers typically correlate (match) the data with a set of pre-determined templates, to find the template that produces the maximal output \citep{cordes09}. Any inaccuracies in the templates compared with the true signal will degrade performance. Matched filters are applied in static fields, and their application becomes problematic for dynamic datasets: the loss in detection performance due to a spatial mismatch of the filter can be compounded by a temporal mismatch.

Recent radio surveys (e.g., FIRST, NVSS, SUMSS) have employed matched filter plus flux limit threshold detectors, typically fitting gaussians with free parameters as the signal filter, and choosing a flux limit based on the estimated noise level of the dataset \citep{becker95,condon98,mauch03}.

Detection of sources in image space can be problematic, firstly because it requires deconvolution of the image prior to detection: over--, under-- and inaccurate cleaning can lead to biased images \citep{condon98}. \citet{perley99} and \citet{rau09} discuss the origin and magnitude of errors in synthesis images. In general, the many-to-one fourier operation performed on visibility data to obtain image data propagates any non-gaussianity in the ($uv$) data to the image plane, producing correlations in the noise between pixels across the field \citep{refregier98}. Imperfect calibration and atmospheric effects yield deviations from gaussianity in visibility data. In addition, images contain structured backgrounds: source sidelobes, gridding artifacts (although, these are minimal for snapshot observations), and, in some cases, confusing sources. The presence of source sidelobes due to data incompleteness confounds identification of true sources. Differencing of temporally adjacent images to detect transient sources is complicated by non-uniform image pixel size and changes in the sidelobe distribution due to the different $uv$-plane sampling. It can also lead to artifacts when subtracting one image with a complicated noise structure, from another with a different noise structure. Finally, removal (flagging) of baselines changes the beam shape, yielding subtraction artifacts in the image. Although data incompleteness and non-gaussian noise also exist in visibility data, in that space we can explicitly consider only the measured data in our detector, and model the deviations from gaussianity: detectors in image space work on the intensity of image pixels, and cannot account for signal that is inferred by the image production process.

We do not mean to suggest that practical and useful source detection cannot be performed in the image plane. \citet{wijnholds08} describe how to propagate errors in visibility space to image space. This type of technique \citep[and others, such as bootstrapping, used by][as a tool to assess radio image fidelity]{athol10} can be used to extend an understanding of the statistical properties of the data into image space. We do, however, regard visibility space as a more natural space for optimal detection, because the data are in a form closer to the original signals collected by the antennas, the covariance structure can be more simply expressed, and no data are inferred through interpolation in the image plane and extrapolation from the $uv$ plane (as is the case with forming an image from incomplete $uv$ data).

An emerging application of general source detection is detection of transient sources. Transient source detection and characterization are key science drivers for many synthesis imaging arrays that are under construction. The Murchison Widefield Array (MWA) and Australian SKA Pathfinder (ASKAP) are two SKA precursor instruments under construction in the Western Australian desert. ASKAP is an array of 36 12-metre dish antennas, currently being constructed at the Murchison Radio-astronomy Observatory (MRO), Western Australia, and operated by CSIRO. The MWA will probe much of the parameter space of interest to the SKA. It is also under construction at MRO, will comprise 512 tile-type antennas with multiple dipole sensors per tile, and is being constructed by a consortium of local and foreign institutions and government agencies. The MWA and ASKAP use fundamentally different hardware for signal reception and processing, and operate in different frequency ranges (ASKAP: 0.7--1.8 GHz, MWA: 80--300 MHz), thereby producing complementary data sets, and pursuing different science goals. The MRO is Australia and New Zealand's candidate site for the core of the SKA.

The MWA and ASKAP are both wide-field instruments, and will both contend with variations in calibration across the field-of-view. As such, they will make use of `field-based' calibration, whereby calibration sources across the field are used to form a model for the antenna beam \citep{kassim07}. The operating frequencies of both instruments also make them subject to atmospheric and ionospheric effects on the signal wavefront. These effects include blurring of the source position, due to phase fluctuations from the troposphere (ASKAP), and source shifting due to differential excess path length to antennas produced by the ionosphere \citep[MWA,][]{mitchell08}. These are some of the challenges faced by wide-field instruments.

Design of transient detectors is relatively new. In general, expected transient source populations will be dominated by weak sources and therefore extraction of sources close to the noise limit will generate the most new and interesting science. Hence, transient detector design is an important field of research. The Allen Telescope Array (ATA) has recently reported the initial development of their slow transient detection pipeline \citep{croft10}. At this preliminary stage, they are matching known catalogues with sources in their fields, and have not found any new convincing transient candidates. Within the ASKAP project, the same fields of sky will be observed periodically to detect slow transients with the VAST survey. An image of the sky will be produced periodically, and these images searched for all sources. Any detections will be added to a searchable database, from which light-curves of objects can be extracted. Transient signals will necessarily show brightness variability over time. The current method being proposed to detect sources is \textit{Duchamp}\footnote[1]{Software and user guide available at \textit{http://www.atnf.csiro.au/people/Matthew.Whiting/Duchamp}}. \textit{Duchamp} uses either a simple thresholding to find sources, or a more sophisticated algorithm based on statistical decision theory. The downside to the method is the use of global parameters, and assumed white gaussian noise properties to define the PDFs. Noise properties can be assumed to be uniform for small fields, but may vary greatly over the field for the large fields-of-view sampled in ASKAP and the MWA. In addition, the technique is not `real-time'. \citet{fridman10} has recently proposed a method for detecting single fast transient events from single-dish datasets, using a cumulative signal method based on statistical decision theory.


Optimal detection of a signal relies on our knowledge of the properties of the signal --- location, shape and amplitude. For signals with unknown parameters (e.g., transient radio sources), the detection performance is governed by our ability to accurately and precisely estimate the parameter values, and accurately model the data likelihood function. It is therefore crucial to understand the fundamental estimation limits of a particular instrument, before proceeding to determine the detection limits \citep[see][for a recent review of fundamental radio imaging limits]{wijnholds08}. In this paper, we describe how to design an optimal source detector with visibility data. We then describe the form of this detector for realistic interferometers, including the effects of imperfect calibration, signal confusion and atmospheric phase fluctuations. As part of this process, and to investigate the detection limits of instruments, we derive fundamental estimation limits for measurement of source parameters with interferometers. In Paper II we explore particular algorithms for source detection with real interferometers and simulated and real datasets, using the estimation limit results and theory from Paper I. We particularly focus on the problem of optimally detecting slow radio transients, although the methods presented here are generally applicable to source detection. As such, the estimation precision results we derive are based on short integrations (8--10s), appropriate for the instruments under consideration, for which a transient detection test is performed at each output of the correlator.

In section \ref{sdt} we introduce statistical decision theory, including methodology for designing an optimal detector. We then introduce the Cramer-Rao lower bound (CRB) on estimation precision, as a metric for evaluating the source parameter measurement precision of interferometers. We then discuss detection of a single point source (section \ref{single_source}), a single point source embedded in a field (section \ref{single_source_field}), and a single transient point source embedded in a field (section \ref{single_source_transient}), with a visibility dataset and thermal noise. In section \ref{calibration_errors} we discuss the effect of calibration errors, source confusion, and atmospheric phase noise on signal estimation and detection, and in section \ref{final_likelihood} present a realistic detector for visibility data.

\section{Statistical decision theory}\label{sdt}
\subsection{Neyman-Pearson test and simple hypothesis testing}\label{np}
Statistical decision theory is the branch of mathematical statistics that describes the detection of signals in noise. Signal detection is underpinned by hypothesis testing: in the binary case, this means deciding between two hypotheses (signal present and signal absent). For a given set of observational data, the likelihood that data was obtained from each hypothesis is calculated, and the ratio of these probabilities is used to compare to a threshold. If the ratio is greater than the threshold, we decide that a signal has been detected. The value of the threshold is set according to the tolerance on the false alarm rate (rate of false positives or misses). This is a likelihood ratio test (LRT), and is applicable for a deterministic signal in known noise. The likelihood is the probability of the data given a set of parameters.

If the null hypothesis (signal absent) is denoted $H_0$, and the alternative hypothesis (signal present) is denoted $H_1$, then 
the hypotheses can be written as:
\begin{eqnarray}
H_1: x[n] &=& s[n] + w[n] \hspace{5mm} (n=1,...,N)\\\nonumber
H_0: x[n] &=& w[n],
\end{eqnarray}
where $s[n]$ is the known deterministic signal we wish to detect, and $w[n]$ is the known noise. Under these two hypotheses, the likelihood ratio test for the dataset $x[n]$ decides a signal is present if,
\begin{equation}
T({\bf{x}}) = \frac{L({\bf{x}};H_1)}{L({\bf{x}};H_0)} > \lambda,
\label{lrt}
\end{equation}
where $T({\bf{x}})$ is the test statistic, $\lambda$ is the threshold, and $L({\bf x})$ is the likelihood function (probability distribution function parametrized by the model parameters). For example, if the signal is a DC level, $A$, in white Gaussian noise (WGN), $N(0,\sigma^2)$, the LRT is:
\begin{equation}
\frac{
\exp\left[-\frac{1}{2\sigma^2}\displaystyle\sum_{n=1}^{N}(x[n]-A)^2\right]}{
\exp\left[-\frac{1}{2\sigma^2}\displaystyle\sum_{n=1}^{N}x^2[n]\right]} > \lambda.
\end{equation}
Taking the logarithm of both sides, and incorporating non-data terms into the threshold, we decide $H_1$ if,
\begin{equation}
T({\bf{x}}) = \frac{1}{N}\displaystyle\sum_{n=1}^{N}x[n] > \lambda^\prime.
\end{equation}
In this simple case, the test statistic is the sample mean, which makes sense intuitively.

We wish to maximise the probability of detection, subject to a given probability of false detection. Mathematically, the probability of detection, $P_D$ is the chance of deciding $H_1$ when the datum is actually drawn from $H_1$, and is given by,
\begin{equation}
P_D \equiv P(H_1;H_1) = \int_{R_1} L({\bf{x}};H_1)d{\bf{x}},
\end{equation}
where $R_1$ is the region of the likelihood function above the threshold. The false positive rate is given by,
\begin{equation}
P_{FA} \equiv P(H_1;H_0) = \int_{R_1} L({\bf{x}};H_0)d{\bf{x}} = \alpha.
\end{equation}
For a given probability of false detection, $\alpha$, the Neyman-Pearson theorem states that the probability of detection, $P_D$, is maximised when we decide $H_1$ according equation \ref{lrt}. Hence, for a given tolerance of false positive signals (chosen according to the detection task and the implications of detecting false positives), the threshold and detection performance are defined. The LRT therefore \textit{yields an optimal detector} ($P_D$ maximised for a given $P_{FA}$).

To design an optimal detector, the likelihood function needs to accurately describe the data. In the following sections, we describe how to accurately model (i) the signal, (ii) the noise and background properties of interferometric data.

\subsection{Generalised Likelihood Ratio Test}
\label{glrt}
The likelihood ratio test described in the previous section can be applied to known deterministic signals with additive known noise. In the case where some parameters are unknown, the values of these parameters need to estimated before proceeding (e.g., source position, flux density). To achieve this, the likelihood function is maximized with respect to each of the unknown parameters, and the LRT is evaluated at the maximum likelihood estimates (MLE). MLEs are asymptotically efficient: for large datasets, they achieve the optimal estimation precision and are unbiased. The hypothesis test is then termed the Generalised Likelihood Ratio Test (GLRT). As the number of unknown parameters increases, the detection performance is degraded. Similarly, as the number of samples increases, the variance on the parameter estimate will decrease, and the estimation will be more precise. In addition to signal parameters that are unknown, a signal model may include additional parameters, in which we are not interested, that need to be estimated. These are nuisance parameters, and further degrade detection performance. The \textit{clairvoyant detector} \citep[Ch. 6]{kay98} assumes perfect knowledge of all parameters, and is useful as a comparison to the real detection performance obtained with the GLRT. The clairvoyant detector is an optimal detector, however the GLRT yields a sub-optimal detector, because estimation of unknown parameters is required. This observation is important when we wish to compare optimal detection performance with the performance we obtain with different detectors we might design.

\subsection{Bayesian approach}\label{bayesian}
In the case where some of the unknown parameters are random (as opposed to deterministic), one can use a Bayesian approach to remove them from the detector. In the Bayesian framework, each datum contains a realization of the random variable, and the likelihood function is conditioned on the parameter. One assigns prior PDFs to the random parameters, and determines the unconditional likelihood functions according to:
\begin{equation}\label{conditional_pdf}
L(\boldsymbol{x};H_1) = \int L(\boldsymbol{x}|\theta;H_1)p(\theta)d\theta,
\end{equation}
where $L(\boldsymbol{x}|\theta;H_1)$ is the conditional PDF, $p(\theta)$ is the prior distribution, and the integration is performed over the random parameter. If the mean value of the random parameter is unknown, one can initially estimate this using MLE. Thus, instead of calculating the MLE of the parameter to completely specify the likelihood function, one can remove it via integration.

\subsection{Detection performance}
As described in section \ref{np}, the detection threshold, $\lambda$ is determined according to the acceptable false positive rate, and balances the probability of detection ($P_D$) with the probability of a false positive $P_{FA}$. It is important to quantify the performance of a detector, for comparison to other detectors. The optimal detection performance, that of the clairvoyant detector, can be calculated and used as a comparison for the realised performance, as an objective means to measure the utility of a detector.

\subsection{Estimation performance: Cramer-Rao lower bound}\label{crb_section}
The complete specification of the likelihood function will require the estimation of some parameters. Imprecise estimation will degrade detection performance. It is useful to have a sense of the ability to estimate the value of a parameter for given a dataset.

To determine the theoretical optimal estimation performance with a given dataset, we can calculate the Cramer-Rao lower bound (CRB) on the precision of parameter estimates. The CRB calculates the precision with which a minimum-variance unbiased estimator could estimate a parameter value, \textit{using the information content of the dataset}. It is computed as the square-root of the corresponding diagonal element of the inverse of the Fisher information matrix (FIM). The ($ij$)th entry of the FIM for a vector ${\boldsymbol{\theta}}$ of unknown parameters is given by:
\begin{equation}
[\boldsymbol{I(\theta)}]_{ij} = -E\left[\frac{\partial^2{\log{L({\bf{x}};{\boldsymbol{\theta}})}}}{\partial{\theta_i}\partial{\theta_j}} \right],
\end{equation}
where $E$ denotes the expectation value. For $N$ independent samples in WGN and complex data, this expression simplifies to \citep{kay93},
\begin{equation}
[\boldsymbol{I(\theta)}]_{ij} = 2{\rm{Re}}\left[\frac{1}{\sigma^2}\displaystyle\sum_{n=1}^N\frac{\partial{\tilde{s}^H[n;{\boldsymbol{\theta}}]}}{\partial{\theta_i}}\frac{\partial{\tilde{s}[n;{\boldsymbol{\theta}}]}}{\partial{\theta_j}}\right].
\label{CRB}
\end{equation}
The CRB is a useful metric because it places a fundamental lower limit on the measurement precision of any parameter. In this work it will be used to gain an understanding of the fundamental limits of an instrument, and how these affect its estimation and detection performance. It has previously been used in astronomy to determine limits on optical astrometry with the WFPC2 camera aboard the Hubble Space Telescope \citep{adorf96}, and with focal plane array bolometers \citep{saklatvala08}.

\section{Detection of sources in visibility data}
We describe here a method for detecting a point source within visibility data. In general, this requires detection of a source of unknown flux density, spectral index, location, arrival time and duration (in the case of a transient), contained within confounding (nuisance) signals within the field. As described in section \ref{glrt}, the method involves maximum likelihood estimation of the unknown parameters, followed by a GLRT to decide the presence of a signal. We begin with the simplest problem of detecting a single point source in an empty field, and then add complexity. It is assumed initially that there are no calibration errors, source confusion or atmospheric effects present in the dataset. These errors will complicate the formulation of the problem, and they will be considered in section \ref{calibration_errors}.

\subsection{Detection of a single point source in an empty field}\label{single_source}
\subsubsection{Estimation of unknown parameters}
Detection of a single point source in visibility data requires estimation of unknown parameters, followed by detection of a complex signal in WGN. We assume we use data from one integration step, $F$ frequency channels and $N$ baselines. The location of the source and its amplitude (spectral flux density) are unknown, and need to be estimated before hypothesis testing. The data are modelled under each hypothesis as:
\begin{eqnarray}
H_1: \tilde{x}[f,n] &=& \tilde{s}[f,n] + \tilde{w}[f,n] \hspace{5mm} (n=1,...,N), (f=1,...,F)\\\nonumber
H_0: \tilde{x}[f,n] &=& \tilde{w}[f,n],
\end{eqnarray}
where the tilde denotes complex quantities. We will write the signal and data as real and imaginary components, under which the noise can be modelled as white gaussian. The signal, $\tilde{s}[f,n]$, is the complex visibility for channel $f$ and baseline $n$, and is given by \citep{thompson04}:
\begin{equation}
\tilde{s}[f,n] = V(u_{fn},v_{fn}) = \iint A(l^\prime,m^\prime)I(l^\prime,m^\prime)\left(\frac{\nu(f)}{\nu_0}\right)^\alpha\exp{\left[-2{\pi}i(u_{fn}l^\prime+v_{fn}m^\prime)\right]}dl^\prime dm^\prime,
\end{equation}
where $A(l^\prime,m^\prime)$ and $I(l^\prime,m^\prime)$ are the antenna response function and source intensity function at sky position $(l^\prime,m^\prime)$, and the spectral dependence is modelled as a power-law with index $\alpha$ and normalized by the base frequency, $\nu_0$. Assuming a point source located at $(l^\prime=l,m^\prime=m)$, the visibility function becomes:
\begin{equation}
V(u_{fn},v_{fn}) = A(l,m)I(l,m)\left(\frac{\nu(f)}{\nu_0}\right)^\alpha\exp{\left[-2{\pi}i(u_{fn}l+v_{fn}m)\right]}.
\end{equation}
The antenna response and source flux density functions are nuisance parameters for detection, and we combine them to form one scaling factor, $B(l,m) = A(l,m)I(l,m)$, which is independent of baseline. Hence, our model is:
\begin{equation}
\tilde{s}[f,n] = V(u_{fn},v_{fn}) = B(l,m)\left(\frac{\nu(f)}{\nu_0}\right)^\alpha\exp{\left[-2{\pi}i(u_{fn}l+v_{fn}m)\right]}.
\label{visibility_signal}
\end{equation}

We form the likelihood function under WGN, assuming initially that the noise variance is known and identical for all baselines and channels. The likelihood function, which is the joint PDF for $F$ channels and $N$ baselines, is given by \citep{kay98}:
\begin{eqnarray}
L(\tilde{\bf{x}};H_1) &=& \prod_{n=1}^N \prod_{f=1}^F \frac{1}{\pi\sigma^2}\exp{\left[-\frac{1}{\sigma^2}(\tilde{x}[f,n]-\tilde{s}[f,n])^*(\tilde{x}[f,n]-\tilde{s}[f,n])\right]}\nonumber\\
&=& \frac{1}{\pi^{NF}\sigma^{2NF}}\exp{\left[-\frac{1}{\sigma^2}(\tilde{\bf{x}}-\tilde{\bf{s}})^H(\tilde{\bf{x}}-\tilde{\bf{s}})\right]}
\end{eqnarray}
where $H$ denotes Hermitian conjugate (complex conjugate tranpose), and the product has been collected in the matrix inner product. Substituting the signal, equation \ref{visibility_signal}, into the likelihood function yields,
\begin{equation}
L(\tilde{\bf{x}};H_1) = \frac{1}{\pi^{FN}\sigma^{2FN}}\exp{\left(-\frac{Z}{\sigma^2}\right)}
\end{equation}
where
\begin{equation}
\begin{split}
Z = \displaystyle\sum_{n=1}^N\displaystyle\sum_{f=1}^F\left(\tilde{x}[f,n]-B(l,m)\left(\frac{\nu(f)}{\nu_0}\right)^\alpha\exp{[-2\pi{i}(u_{fn}l+v_{fn}m)]}\right)^*\\
\times\left(\tilde{x}[f,n]-B(l,m)\left(\frac{\nu(f)}{\nu_0}\right)^\alpha\exp{[-2\pi{i}(u_{fn}l+v_{fn}m)]}\right).
\end{split}
\end{equation}
To obtain the values of the unknown parameters, $(B,\alpha,l,m)$, the maximum likelihood estimates of these parameters, $(\hat{B},\hat{\alpha},\hat{l},\hat{m})$, are determined. Methods for determining these estimates are presented in Paper II. Here, we explore the precision with which these parameters can be estimated for a real instrument using the CRB formalism described in Section \ref{crb_section}.

\subsubsection{Estimation precision for a point source}
For the signal, equation \ref{visibility_signal}, the FIM for parameters ($l,m,\alpha,B$) is given by;
\begin{eqnarray}
[\boldsymbol{I(\theta)}] &=& \frac{2}{\sigma^2}
\begin{pmatrix}
4\pi^2{B}^2 I_{u^2} & 4\pi^2{B}^2 I_{uv} & 0 & 0 \\
4\pi^2{B}^2 I_{uv} 
& 4\pi^2{B}^2 I_{v^2} & 0 & 0 \\
0 & 0 & NB^2I_{\nu^2} & NB\,
I_{\nu} \\
0 & 0 & NB \, I_{\nu} & N I_{1}
\label{matrix}
\end{pmatrix}, \\
\hbox{where } &I_{u^2}& = \displaystyle\sum_{n=1}^N \displaystyle\sum_{f=1}^F  u_{fn}^2\left(\frac{\nu(f)}{\nu_0}\right)^{2\alpha},  \\
&I_{uv}& = \displaystyle\sum_{n=1}^N \displaystyle\sum_{f=1}^F  u_{fn}v_{fn}\left(\frac{\nu(f)}{\nu_0}\right)^{2\alpha},  \\
&I_{v^2}& = \displaystyle\sum_{n=1}^N \displaystyle\sum_{f=1}^F v_{fn}^2\left(\frac{\nu(f)}{\nu_0}\right)^{2\alpha}, \\
&I_{\nu^2}& = \displaystyle\sum_{f=1}^F \left(\frac{\nu(f)}{\nu_0}\right)^{2\alpha} [\log{(\nu(f)/\nu_0)}]^2, \\
&I_{\nu}& = \displaystyle\sum_{f=1}^F \left(\frac{\nu(f)}{\nu_0}\right)^{2\alpha} \log{(\nu(f)/\nu_0)},  \\
\hbox{and } &I_{1}& = \displaystyle\sum_{f=1}^F \left(\frac{\nu(f)}{\nu_0}\right)^{2\alpha}. 
\end{eqnarray}
Interestingly, the FIM does not depend directly on the source location (only indirectly through the antenna response function). This is because it is only the phase difference between antennas that is important. Instead, the baseline projections $(uv)$ weight the information in each element of the summations. Intuitively this is because the longer baselines are sensitive to smaller changes in the source position, and therefore are weighted more highly in the information measure. Because the intensity scaling, $B$, is a linear multiplier for each signal, the information carried in the data about it scales directly with the number of baselines, and the noise variance. There is no covariance between the position parameters and the signal amplitude parameters. Therefore, any prior information on the values of these parameters, will not affect one's ability to estimate the other parameters. Conversely, any factor that degrades the estimation of one group will not affect the other group.

Inverting the FIM yields the following lower bounds on the precision of the parameter estimates:
\begin{eqnarray}
\Delta{l} &\geq& \frac{\sigma I_{v^2}^{1/2}}{2\sqrt{2}\pi{B}} 
\left[ I_{v^2} I_{u^2} - \left(I_{uv} \right)^2 
\right]^{-1/2} \label{l_precision}\\
\Delta{m} &\geq& \frac{\sigma I_{u^2}^{1/2}}{2\sqrt{2}\pi{B}} 
\left[ I_{v^2} I_{u^2} - \left(I_{uv}\right)^2 \right]^{-1/2} \\
\Delta{\alpha} &\geq& \frac{\sigma I_1^{1/2} }{\sqrt{2N}B} \left[
 I_{\nu^2}  I_1 -  \left( I_{\nu} \right)^2 \right]^{-1/2} \\
\Delta{B} &\geq& \frac{\sigma I_{\nu^2}^{1/2} }{\sqrt{2N}} \left[
 I_{\nu^2} I_1 - \left( I_{\nu} \right)^2 \right]^{-1/2}\end{eqnarray}
Note that the sky positions here are direction cosines, and are defined in radians for small angles, and $B$ and $\sigma$ have the same units (i.e., Jy). This is a general expression for the theoretical maximum precision (estimation performance) on the position, amplitude and spectral index of a source in visibility data at $(l,m)$, using $F$ frequency channels and $N$ baselines. The noise, $\sigma$, is thermal noise per channel and baseline. These expressions exclude any systematic effects. They are the lower bounds on estimation of these parameters for an estimator that uses all of the available information in an unbiased manner.

Equation \ref{l_precision} suggests that a centrally concentrated array will have poorer astrometric precision than an array that is not centrally concentrated but has the same number of baselines and the same maximum baseline (i.e., one with uniform $uv$ coverage). In conventional radio astronomy terminology, a naturally-weighted synthesized beam from a concentrated array will have a broader beam than a uniform $uv$ array. These results agree with, and quantify, our intuitive expectations.

\begin{sidewaysfigure}
\subfigure{\includegraphics[scale=0.4]{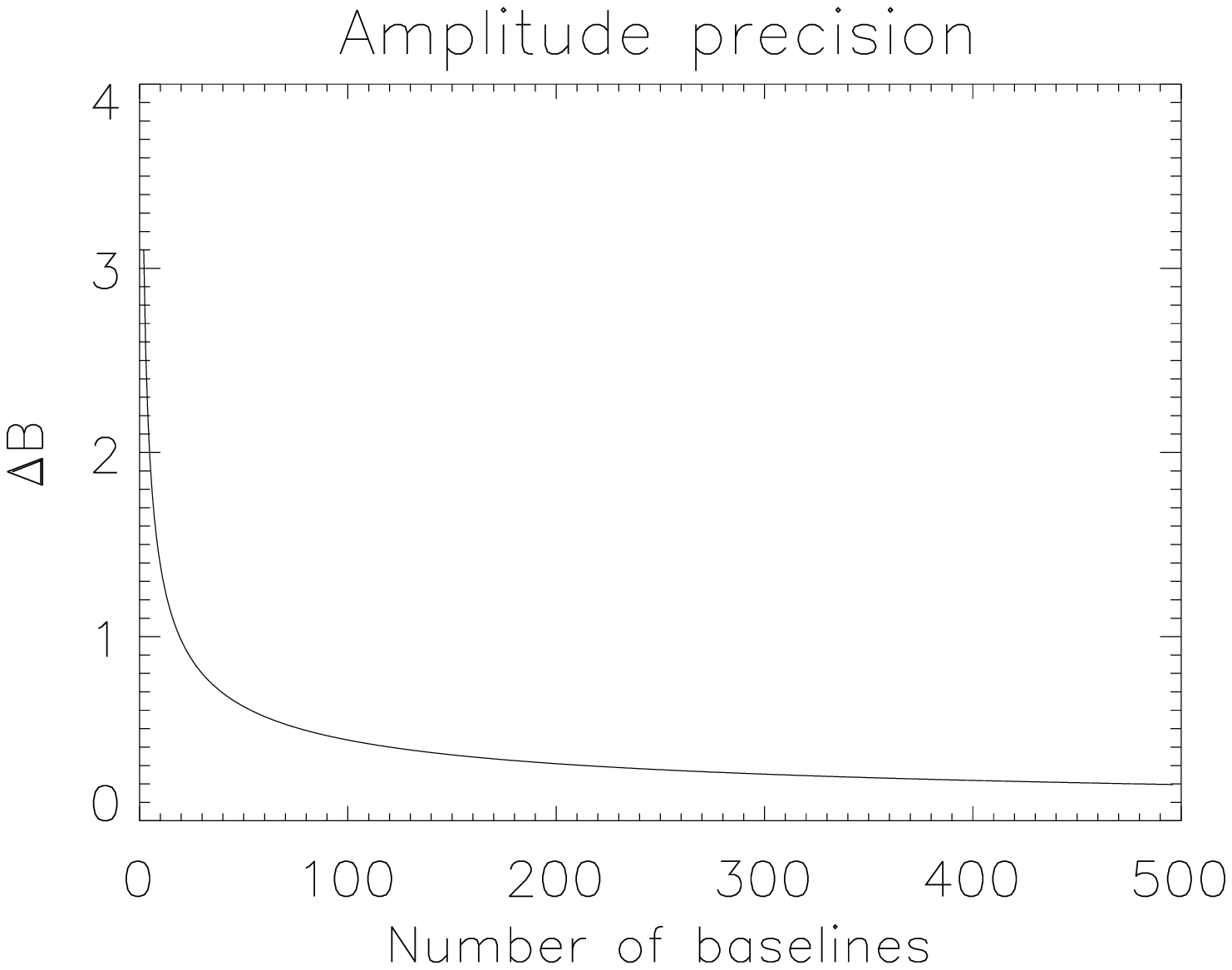}}
\subfigure{\includegraphics[scale=0.4]{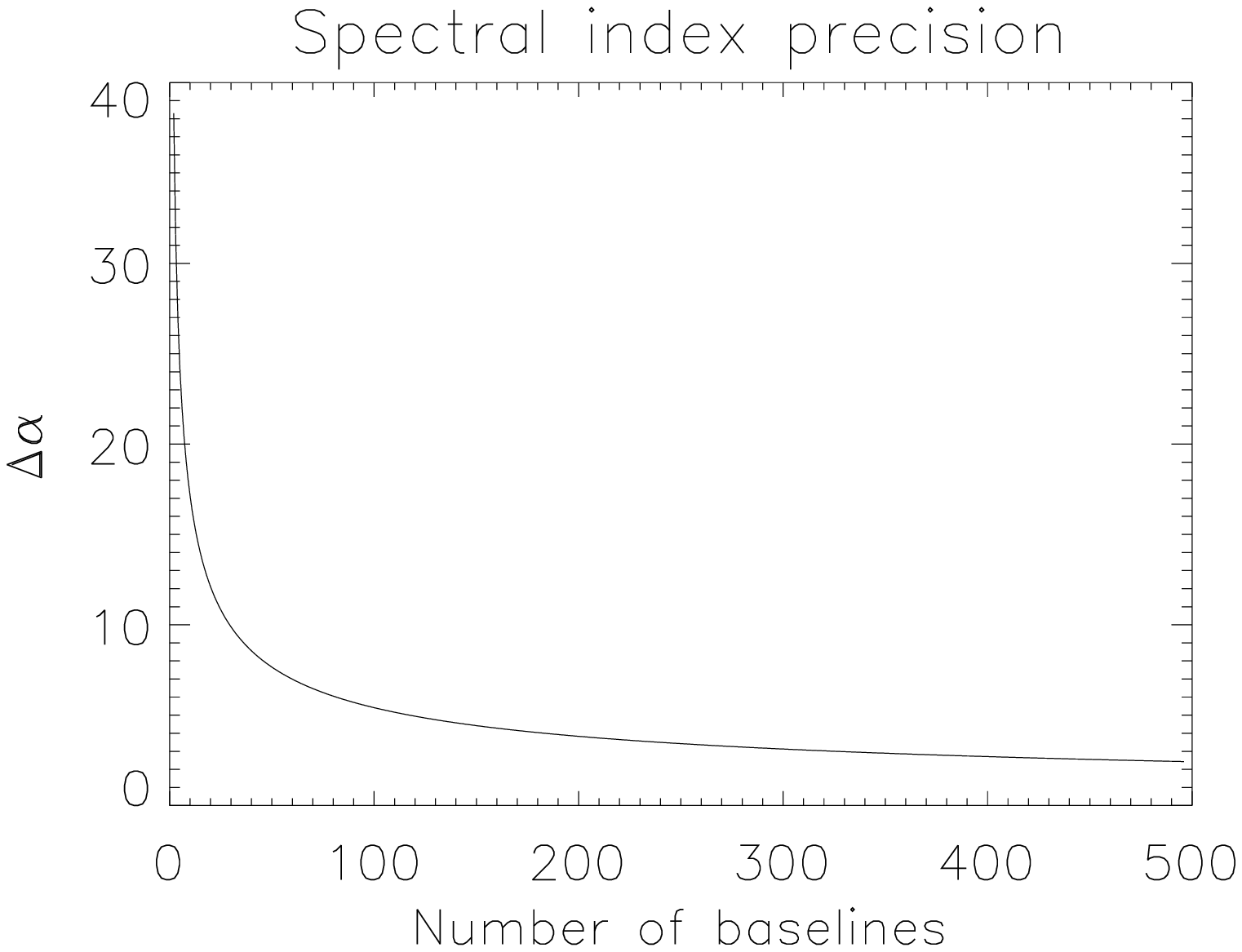}}
\subfigure{\includegraphics[scale=0.4]{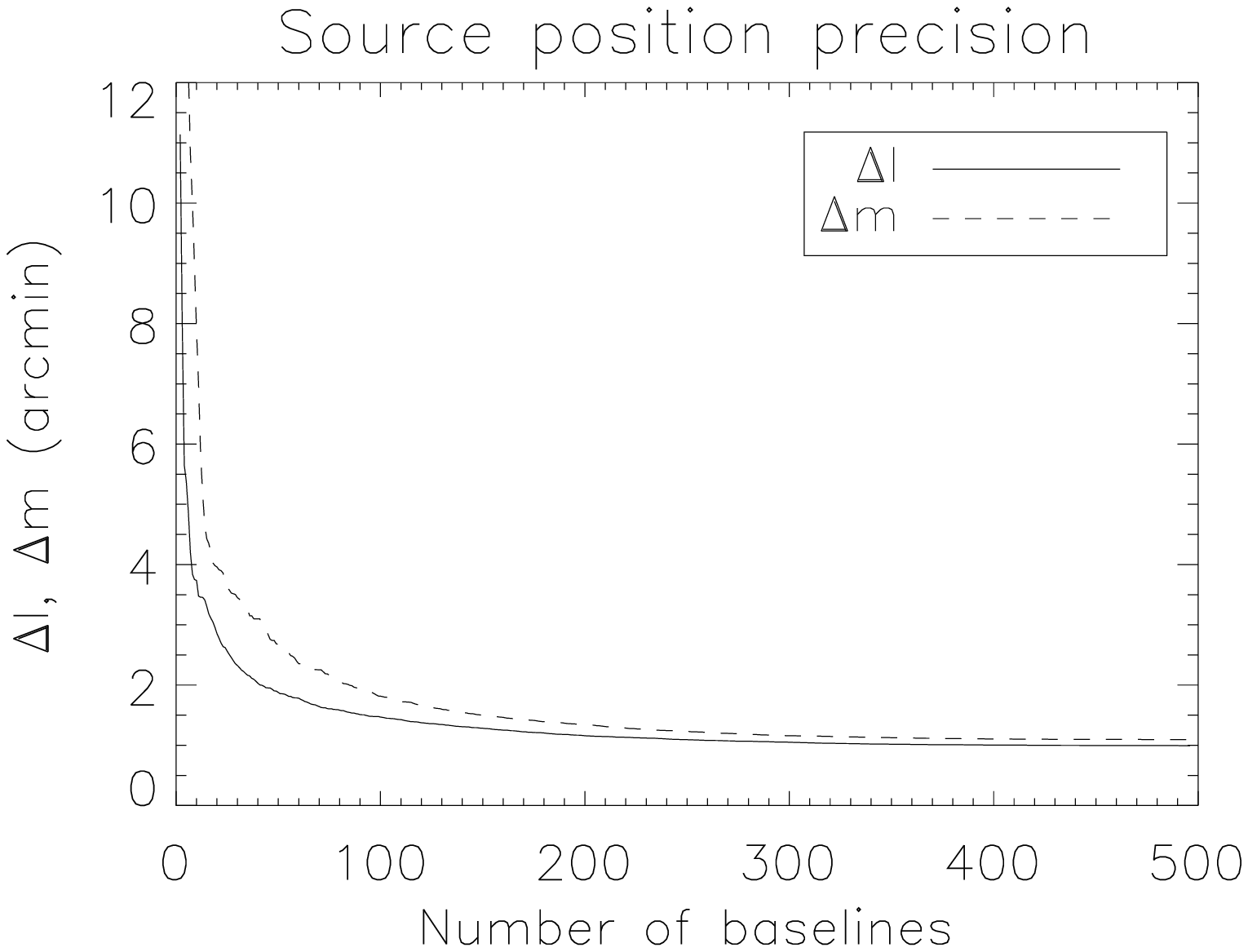}}\\
\subfigure{\includegraphics[scale=0.4]{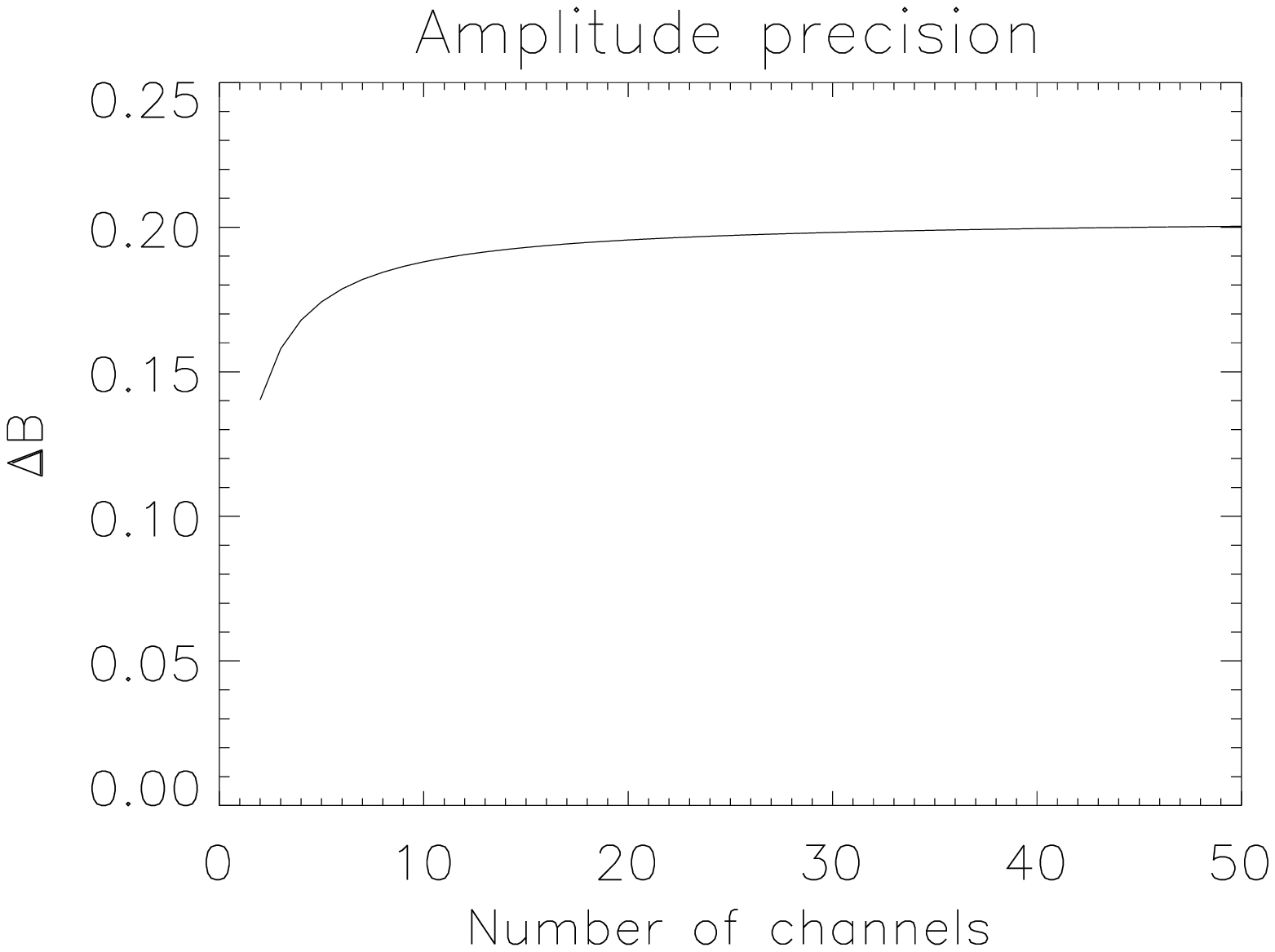}}
\subfigure{\includegraphics[scale=0.4]{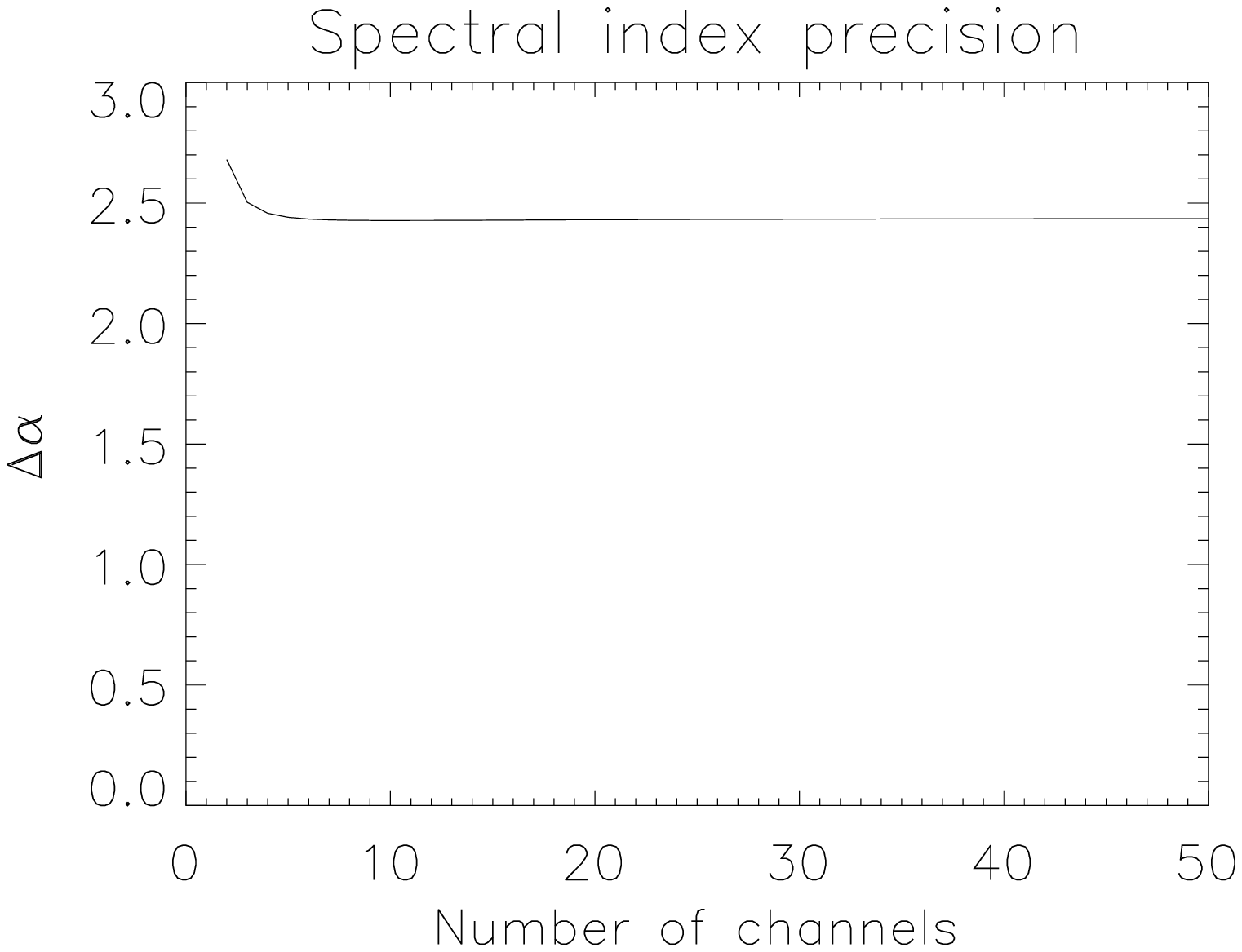}}
\subfigure{\includegraphics[scale=0.4]{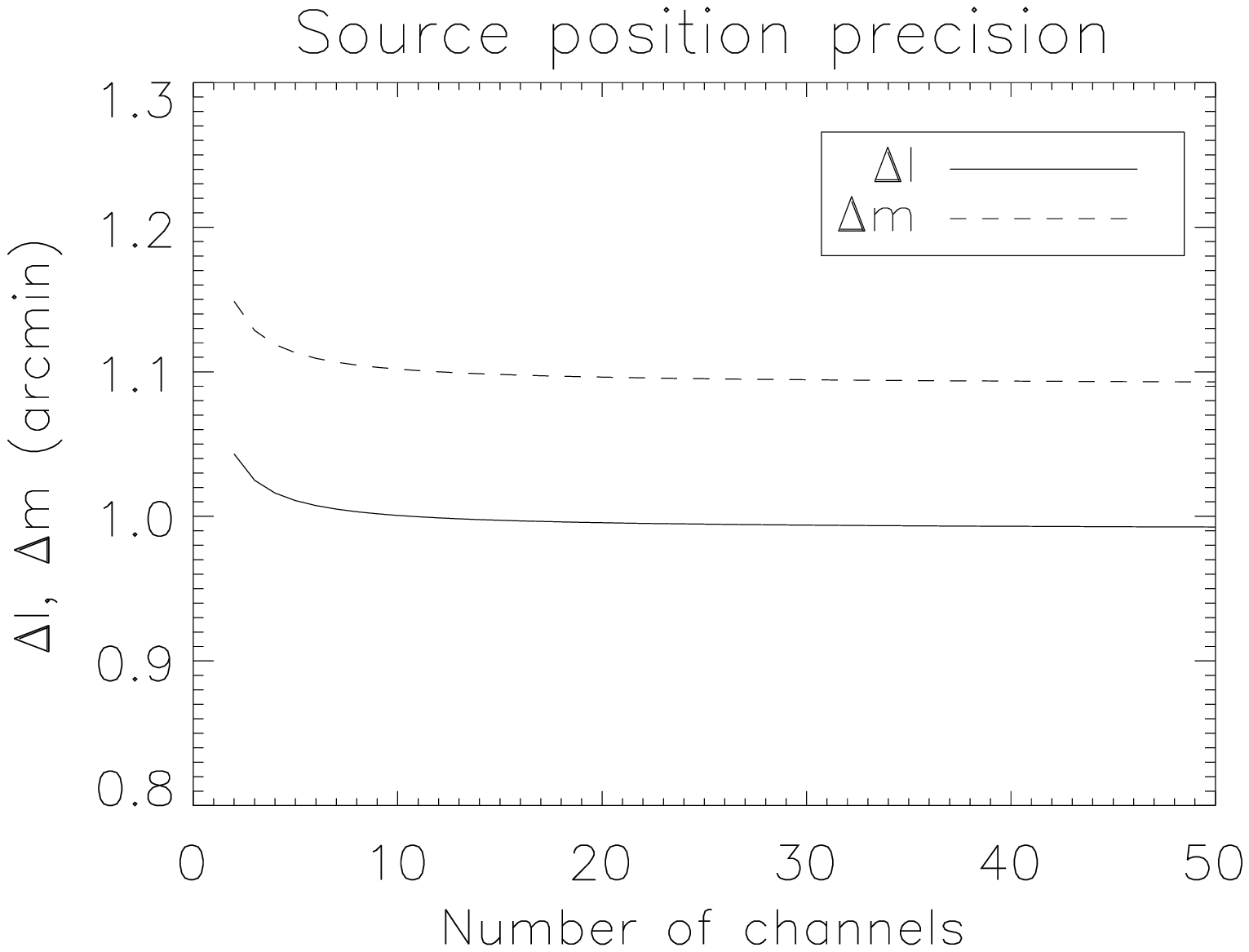}}
\vspace{1cm}
\caption{(Upper) Bounds on unknown parameters as a function of number of baselines, where the baselines have been ordered descending in length ($B=0.7$ Jy, $\sigma=15$ Jy/channel/baseline, $\alpha=0$, $\Delta{t}=8$s, $F=24$, $\Delta{\nu}=30.72$ MHz). (Lower) Bounds as a function of number of frequency channels, $F$, using all 32 antennas, and the same total bandwidth, $\Delta\nu=30.72$ MHz.}
\label{precision_baselines}
\end{sidewaysfigure}

\begin{figure}
\subfigure{\includegraphics[scale=0.16]{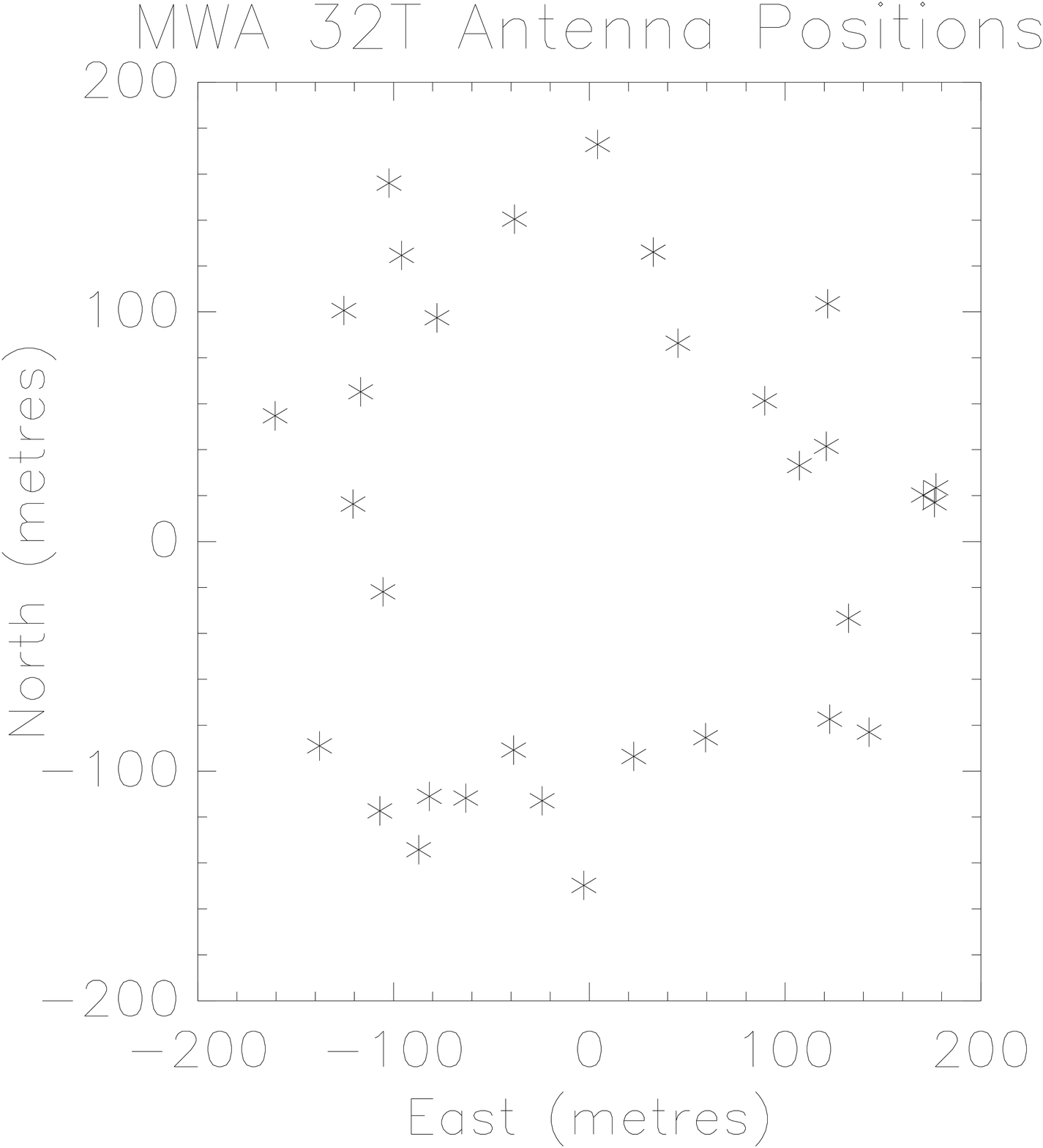}}
\hspace{0.3cm}
\subfigure{\includegraphics[scale=0.16]{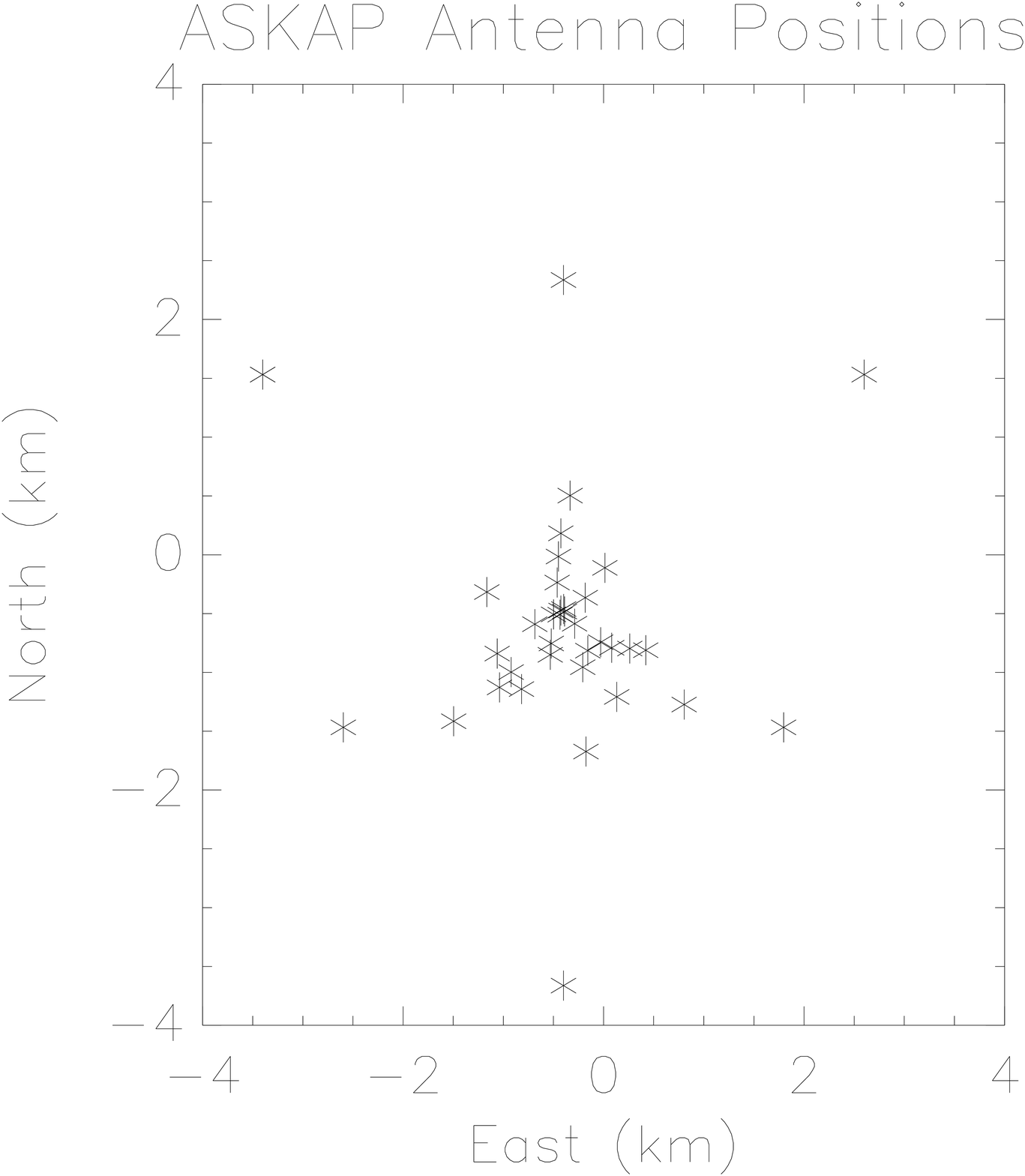}}
\vspace{1.5cm}
\caption{Antenna positions for the MWA 32T telescope (left) and ASKAP telescope (right).}
\label{antenna_locations}
\end{figure}

The covariance (non-zero off-diagonal elements of the FIM) between the signal amplitude and spectral index degrade the estimation performance for each parameter individually. If the spectral index is known, and for $\alpha=0$, the precision on the signal amplitude is given by $\Delta{B} \geq \sigma/\sqrt{2FN}$ i.e., it is proportional to the noise, integrated over all antennas and the total bandwidth. Introducing uncertainty in the value of the spectral index degrades the estimation of the signal amplitude, and vice versa. Figure \ref{precision_baselines} shows estimation precision for the 32-tile system (32T) of the MWA, where the phase centre has been set at the zenith. The MWA 32T has a linear extent of $\sim$330m, and a synthesized beam at 150 MHz of $\theta_{\rm syn}\sim$25 arcmin. Figure \ref{antenna_locations} displays the antenna positions for the MWA 32T and ASKAP telescopes.
Assuming a system equivalent flux density of 10,000 Jy, 1.28 MHz channels and an 8 second integration, the noise in each of 24 channels is $\sim$15 Jy. In all cases, the source has a constant amplitude of $B=0.7$ Jy, translating to an integrated signal-to-noise ratio, S/N=5, over 30.72 MHz bandwidth (center frequency of 153 MHz) and using all 32 antennas. 
The upper figures display bounds as a function of the number of baselines, where the baselines have been ordered descending in length (i.e., the longest baselines are counted first), and for 24 frequency channels. The lower panels display bounds as a function of the number of spectral channels used for a fixed total bandwidth of 30.72 MHz, constant source amplitude, and using all antennas. The following observations can be made:
\begin{itemize}
\item Inclusion of the shortest baselines does affect estimation performance, due to the increase in sensitivity obtained by including additional antennas. This is important when considering removing short baselines to reduce the impact of diffuse emission on your signal;
\item The Fisher information on the source amplitude and spectral index are independent of the distribution of antennas in the array: they depend on the number of antennas, frequency channels and bandwidth, and have a $1/\sqrt{N}$ dependence on number of baselines;
\item Increasing the number of frequency channels degrades amplitude estimation performance, but improves spectral index estimation performance, due to the covariance of these parameters;
\item The position parameters have the same functional form, and improvement in one generally implies improvement in the other (unless the improvement is due to increased interferometer extent along only one axis);
\item In general, the uncertainty in the value of the spectral index is large compared with typical values, and this poor estimation is due to the high noise (short integration time) in this example;
\item The bounds for $m$ are slightly higher than for $l$ due to the reduced spatial extent of the MWA 32T array in the $y$-direction (elongated in the $x$-direction), but both are $\sim$1 arcminute for S/N=5;
\item The CRBs on position estimates are much smaller than the synthesized beam of the telescope. This is because the CRB represents the maximal precision, and thus represents the performance of the optimal estimator (or deconvolution algorithm), if it exists (which it may not). The synthesized beam, or half-power beam width, is a coarse measure that only considers the contribution from the longest baseline in the array, without reference to the spatial location information carried by the shorter baselines.
\end{itemize}

These results are for a ``perfect" interferometer, with no calibration errors and no systematic bias. Once additional errors are taken into account, the effective noise term ($\sigma$) will increase, and the estimation performance will be degraded. This will be closer to the real situation in an interferometer, but will still omit any errors introduced by systematic bias i.e., the CRB is applicable only for unbiased estimation.

\subsection{Inclusion of other sources in the field}\label{single_source_field}
Up to this point, we have considered the field to contain a single source, with unknown position and amplitude. We now generalise these results to the more realistic situation where the field contains many other static sources. From the perspective of transient detection, these static sources are nuisance signals that need to be removed. Removal of these signals might involve signal subtraction, or modelling, and this is a large field of current research, in itself. Algorithms that account for static sources will be explored in Paper II.

These sources are modelled in the visibility data, and included in the GLRT as known parameters, \textit{viz};
\begin{equation}
\frac{p(\tilde{\bf{x}};\hat{B},\hat{\alpha},\hat{l},\hat{m},{\boldsymbol{\theta}},H_1)}{p(\tilde{\bf{x}};H_0)} > \lambda,
\end{equation}
where ${\boldsymbol{\theta}}$ is a vector that describes the parameters of the nuisance sources. The ML estimation of the source position and amplitude can proceed under this scheme as described above. 

For $K$ known static sources, and one unknown source, the signal is given by the sum of the complex visibilities for each source,
\begin{equation}
\tilde{s}[f,n] = B(l,m)\left(\frac{\nu(f)}{\nu_0}\right)^{\alpha}\exp{\left[-2{\pi}i(u_{fn}l+v_{fn}m)\right]} + \displaystyle\sum_{k=1}^K B_k\exp{\left[-2{\pi}i(u_{fn}l_k+v_{fn}m_k)\right]}.
\end{equation}

\subsection{Unknown signal arrival time}\label{single_source_transient}
Transient sources, by their nature, appear at unknown times (for non-periodic sources) and remain visible for unknown durations. Including these parameters in the modelling necessarily requires extending the discussion to multiple integration steps. This has the benefit of increasing the number of data samples, thereby improving parameter estimation, but is complicated by the evolution of $(uv)$ as the Earth rotates.

The ML estimation and GLRT detection test scheme described above, applied at each integration timestep (output of the correlator), naturally allow for appearance of a signal at a given timepoint. No signal-present will produce amplitude estimates within the noise level, and position estimates within the CRB of the phase centre. Appearance of a (sufficiently strong) signal will produce non-zero estimates and the test statistic will exceed the detection threshold. At this point, it is statistically advantageous to include all previous timesteps when a signal has been present in the ML estimation of parameters. The position of the source $(l,m)$ and amplitude, $B(l,m)(\nu(f)/\nu_0)^\alpha$, are constant over time. The GLRT then becomes:
\begin{equation}
L(\tilde{\bf{x}},\hat{B},\hat{\alpha},\hat{l},\hat{m},{\boldsymbol{\theta}};H_1) = \frac{1}{(\pi\sigma^2)^{NFT}}
\exp{\left\{-\frac{1}{\sigma^2}\left[\displaystyle\sum_{t=1}^T\displaystyle\sum_{n=1}^N \displaystyle\sum_{F=1}^F Z_{fnt}\right]\right\}},
\end{equation}
where
\tiny
\begin{equation}
\begin{split}
Z_{fnt} = &\left(\tilde{x}[f,n,t]-B(l,m)\left(\frac{\nu(f)}{\nu_0}\right)^{\alpha}\exp{[-2\pi{i}(u_{fnt}l+v_{fnt}m)]}-\displaystyle\sum_{k=1}^{K} B_k\exp{[-2{\pi}i(u_{fnt}l_k+v_{fnt}m_k)]}\right)^*\\
&\times \left(\tilde{x}[f,n,t]-B(l,m)\left(\frac{\nu(f)}{\nu_0}\right)^{\alpha}\exp{[-2\pi{i}(u_{fnt}l+v_{fnt}m)]}-\displaystyle\sum_{k=1}^{K} B_k\exp{[-2{\pi}i(u_{fnt}l_k+v_{fnt}m_k)]}\right),
\end{split}
\end{equation}
\normalsize
and $t=(1,...,T)$ denotes the timestep, $T$ timesteps have occurred since the initial detection of the transient signal, and the summation contains the contribution from the known, static sources within the field. The baseline projections are now functions of time as the Earth rotates, and are completely specified. Estimation of the transient source position and amplitude is extended to include all $T$ timesteps, effectively reducing the estimation uncertainty by a factor of ${\sim}1/\sqrt{T}$. 
Hence, at each integration timestep, after initial detection of a transient signal, two processes would occur. Firstly, the total dataset (since detection) would be used to estimate the values of the source parameters. Then, these ML estimates would be used in a GLRT, \textit{evaluated at the current timestep alone}. This would allow a decision to be made about the continued presence of the source at that timestep. If all of the data were used for the detection, previous detections of the signal (in previous timesteps) would dilute the presence of the signal in the current timestep. In other words, if the signal has disappeared, it will be more difficult to detect this if all of the previous signal-present information is used.

\section{Other sources of uncertainty}
Up to this point we have considered only thermal noise in visibility data, however this is a simplification of reality. Amplitude and phase calibration errors, background confusing sources and atmospheric/ionospheric effects on the signal wavefront alter the noise properties of the visibility data. Note that here we use the term `noise' in the general sense, including statistical noise, system noise and unresolved background. Accurate characterization of these effects is required to design an optimal detector. As well as designing an optimal detector, these effects introduce additional uncertainty to the modelling and will degrade the estimation and detection performance. In this section we use the Cramer-Rao bound to determine the precision on measuring calibration gain parameters, and include this additional uncertainty in the likelihood function describing the data.

To include the additional uncertainty in the likelihood function, we add a term to the thermal noise that quantifies the uncertainty for each baseline and channel. In general, this is a covariance matrix, $\boldsymbol{C}_c$, where non-zero off-diagonal terms quantify baseline-baseline covariances. The likelihood function under the signal-present hypothesis becomes;
\begin{equation}
L(\tilde{\bf{x}};H_1) = \frac{1}{\pi^{FN}\det(\boldsymbol{C}_c+\sigma^{2}\boldsymbol{I})}
\exp{\left[-(\tilde{\bf{x}}-\tilde{\bf{s}})^H(\boldsymbol{C}_c+\sigma^{2}\boldsymbol{I})^{-1}(\tilde{\bf{x}}-\tilde{\bf{s}})\right]}.
\label{cal_like}
\end{equation}
Multiplying the data model vector by the inverse of the covariance matrix prewhitens the data (removes the correlations between baselines, and weights each baseline according to the amount of information available about it).

\subsection{Calibration errors, confusion and atmospheric phase noise}\label{calibration_errors}
\citet{liu10} have recently described errors introduced by different calibration techniques, and extend earlier work by \citet{cornwell81}, \citet{cornwell89} and \citet{wieringa92}. Calibration errors can be classified into two types; (1) systematic bias, due to a biased calibration estimation method; and (2) estimation uncertainty (imprecision), due to limited information (with the Cramer-Rao bound as the lower limit). Systematic bias will shift the position of sources, but may not increase the model uncertainty. Bias on an antenna-by-antenna basis will blur the position of signals. We will consider unbiased estimation precision, and represent the amplitude and phase calibration errors as additive parameters with zero mean and known covariance.

Low-resolution instruments may suffer from source confusion, whereby the density of background sources is sufficient to produce overlapping sources in the image plane through the source primary signal and sidelobes. Confusion-limited instruments have a natural detection limit that corresponds to the confusion level, as opposed to the thermal noise level, which may be lower. The confusing signal is structured and behaves differently to thermal noise, because it corresponds to real signals. The MWA 32T and 512T are confusion-limited instruments.

In general, the ASKAP instrument will not be confusion-limited, due to its higher angular resolution compared with the MWA 32T and higher operating frequencies (this may not be the case for long integrations, but will be when considering each integration timestep independently). However, the higher frequencies are subject to tropospheric fluctuations, yielding visibilities that include atmospheric phase noise. This noise acts to blur the position of the source. The phase noise is a function of the baseline length, $d$, and its variance can be modelled by \citep{thompson04}:
\begin{equation}
\sigma_{\rm atmos}^2 = \frac{4\pi^2{a}^2{d}^{2\beta}}{\lambda^2},
\end{equation}
where $\beta$ is the index of the structure function describing the fluctuations ($\beta$=5/6 for a Kolmogorov spectrum), and $a$ is a scaling factor. The phase noise is largest for the longest baselines. A typical value for $a$ of 10$^{-6}$ corresponds to an rms phase noise of 2.5 degrees for the longest ASKAP baseline.

Confusion will increase the effective noise level in the visibilities, and the rms confusing signal can be added in quadrature to the thermal noise. Atmospheric phase noise introduces additional uncertainty in the phase, and therefore in the argument of the trigonometric functions describing the signal. Before their inclusion into the detector, we derive the impact of calibration on source parameter estimation precision.

There are two steps in determining the effect of calibration uncertainty on estimating the parameters of a source. The first is to determine how precisely calibration can be performed given a set of calibrators in the field. The second is to include this uncertainty in the overall system noise when estimating the source parameters.

\subsubsection{Form of covariance matrix}

Primary (amplitude) calibration is achieved by observing a source of known flux density, typically at the phase centre, and adjusting the antenna-based gains to yield the known flux density as an output. Secondary (phase) calibration can be performed in two ways: (1) observation of a bright point source at the phase centre, and adjustment of the antenna-based phases to be identically zero, and (2) self-calibration, using sources available in the field to produce a consistent phase solution. For single-dish instruments and interferometers with a small field-of-view, the former technique yields adequate results. For instruments with large fields-of-view, where multiple phase solutions are required (variation in calibration across the field), self-calibration will produce more accurate results at the edge of the field. Observations at frequencies below $\sim$300 MHz have the additional complication of propagation delays introduced by the ionosphere. This delay shifts the position of sources in the sky, but does not blur the image. In this case, one forms a simple time-dependent model for the ionospheric phase screen from sources in the field-of-view, and performs a `field-based' calibration \citep{kassim07}. This requires a high cadence of phase calibration to be performed, and to be practical, necessitates the field-based calibration method. At higher frequencies, phase noise caused by the neutral troposphere causes a blurring of the source position.

An adequate distribution of secondary calibrators across the field should produce unbiased phase solutions, with the spatial variation accounted for in the solutions. The phase errors in this case will reduce to measurement errors based on the number and strength of the sources available. An inadequate density of sources may lead to large uncertainty on the phase calibration. Primary (bandpass) calibration typically employs a very strong source, occurs relatively infrequently ($\sim$hours), and is performed for each frequency channel. The high source signal-to-noise ratio will yield high precision on the calibration. The field-based calibration, however, is subject to short timescale atmospheric fluctuations. For these instruments it will employ sources of varying strengths, occur frequently ($\sim$10s), and will involve simultaneous solution for all antennas across the whole bandwidth. Here we will consider the effect of field-based calibration on source estimation.

The covariance matrix element for baselines $i,j$ is given by;
\begin{equation}\label{covariance}
C_{ij} = E[(\tilde{x}_i-\tilde{\bar{x}}_i)(\tilde{x}_j-\tilde{\bar{x}}_j)],
\end{equation}
where the expectation value is taken over multiple realisations of the calibration. With an ionospheric model and a given density of static field sources, the covariance matrix can be approximated analytically. For more complex and realistic distributions, simulations can be used to calculate the covariance matrix empirically. Real datasets may also be used to quantify the covariance matrix: each independent integration in an observation of a static field can be used to approximate an independent noise realization, and the covariance matrix estimated using equation \ref{covariance}. For the purposes of this paper, and to demonstrate the magnitude and impact of these errors on source estimation and detection, we form an approximate analytical covariance matrix based on the theoretical precision with which calibration can be performed, and known expressions for the magnitude and distribution of uncertainty introduced by confusion and atmospheric phase noise.


To approximate the form of the covariance matrix, $\boldsymbol{C}_c$, we consider the measurement errors for amplitude and phase calibration for a given antenna, and use error propagation to express the variance for a given baseline. Errors in radio astronomy are typically antenna-based, however the covariance matrix we require needs to describe uncertainty on a baseline basis (since these are the data we measure). Note that the formulation of the problem and the error propagation take into account the connectivity of antennas: i.e., an error on one antenna will propagate through all of the baselines it forms with every other antenna.

\subsubsection{Cramer-Rao bounds on estimation of gain parameters}
We begin by calculating the theoretical optimal precision with which the amplitude and phase calibration can be measured for a single antenna, for an $M$-antenna interferometer and $N_c$ calibrators, with positions ($l_{N_c},m_{N_c}$) and flux densities, $B_{N_c}(\nu(f)/\nu_0)^{\alpha_{N_c}}$. We write the complex gain for baseline $n$ comprising antennas $\beta$ and $\gamma$, as:
\begin{equation}
\tilde{G}_n = \tilde{G}_\beta \tilde{G}_\gamma = \frac{1}{b_\beta b_\gamma} \exp{2\pi i(\phi_\beta-\phi_\gamma)},
\end{equation}
where $b_\beta$ and $\phi_\beta$ are the amplitude and phase gain parameters for antenna, $\beta$.
The joint PDF for all of the baselines is proportional to:
\begin{equation}
L(\tilde{\boldsymbol{x}}) \propto \exp{\left[-\frac{1}{2\sigma^2} \displaystyle\sum_{f=1}^F \displaystyle\sum_{\beta=1}^M \displaystyle\sum_{\gamma{\neq}\beta}^M Z_{\beta\gamma}^HZ_{\beta\gamma}\right]}
\end{equation}
where
\begin{equation}
Z_{\beta\gamma} = \tilde{x}_{\beta\gamma} - \displaystyle\sum_{i=1}^{N_c} B_i\left(\frac{\nu(f)}{\nu_0}\right)^{\alpha_i}b_\beta{b_\gamma}\exp{-2\pi i(u_{f\beta\gamma}l_i+v_{f\beta\gamma}m_i+\phi_\beta-\phi_\gamma)},
\end{equation}
and there are $M$ antennas.

The Fisher Information Matrix is a $2M\times 2M$ matrix to estimate all of the $b$ and $\phi$ parameters. There are no covariances between the $b$ and $\phi$ parameters, so the FIM is equivalent to two $M\times M$ matrices. 
Therefore, there are two FIMs to invert, FIM$_b$ and FIM$_\phi$. Constructing FIM$_\phi$ yields a singular matrix, due to the phases being relative quantities. Typically, the phase gain for one antenna is set to zero, and the others are defined relative to this. Hence, we set $\phi_1=0$, and remove this parameter from the estimation (it is assumed completely specified). Therefore FIM$_\phi$ becomes a $(M-1) \times (M-1)$ matrix.

We derive the CRBs in Appendix \ref{appendix_crb} and present the solutions here. The general solutions for the baseline precision $N_c$ calibrators and $M$ antennas are:
\begin{eqnarray}
\Delta{b}_{\alpha\beta} &\geq& \frac{\sigma}{\sqrt{2}} \left[ \displaystyle\sum_{f=1}^F\left(\displaystyle\sum_{i=1}^{N_c} B_i^2\left(\frac{\nu(f)}{\nu_0}\right)^{2\alpha_i}  
\right. \right. \nonumber \\ &\null& \qquad \quad \left. \left.
+ \displaystyle\sum_{i=1}^{N_c}\displaystyle\sum_{j\neq{i}}^{N_c}B_iB_j\left(\frac{\nu(f)}{\nu_0}\right)^{\alpha_i+\alpha_j} \cos{2\pi(u_{f\alpha\beta}(l_j-l_i)+v_{f\alpha\beta}(m_j-m_i))}\right) \right]^{-1/2} \label{b_precision}\\
%
\Delta\phi_{\alpha\beta} &\geq& \frac{\sigma}{\sqrt{2(2\pi)^{2}}}
\sqrt{\varepsilon_{ab...z\neq\alpha\neq\beta}A_{1,a}A_{2,b}...A_{M-1,z}}
\Bigg{/}
\sqrt{\varepsilon_{ab...z}A_{1,a}A_{2,b}...A_{M-1,z}},
\label{calibration_precision}
\end{eqnarray}
where $\varepsilon$ is the Levi-Civita permutation symbol, $(a,b,...,z)\in{[1,M-1]}$, there are implicit summations over all indices, and $A_{a,b}$ are the FIM matrix elements, and are given by,
\begin{equation}
A_{a,b}=\left\{\begin{array}{cl}
	b_a^2\displaystyle\sum_{k\neq{a}}^{M-1}b_k^2X_{ak}, & a=b \\
	-b_a^2b_b^2X_{ab}, & a\neq{b}
	   \end{array}\right.,
\end{equation}
and $X_{ab}$ is given by:
\begin{equation}
\displaystyle\sum_{f=1}^F \left[\displaystyle\sum_{i=1}^{N_c} B_i^2\left(\frac{\nu(f)}{\nu_0}\right)^{2\alpha_i} + \displaystyle\sum_{i=1}^{N_c}\displaystyle\sum_{j\neq{i}}^{N_c}B_iB_j\left(\frac{\nu(f)}{\nu_0}\right)^{\alpha_i+\alpha_j} \cos{2\pi(u_{fab}(l_j-l_i)+v_{fab}(m_j-m_i))}\right].
\label{Xeqn}
\end{equation}
The expression for the phase uncertainty is not easy to implement, and in practise, it is far simpler to form the FIM and invert numerically, and extract the baseline-based uncertainties using error propagation on the inverse FIM elements.

Equations \ref{b_precision}-\ref{calibration_precision} are the baseline-based gain precision limits, and we have used error propagation (including covariances) from the antenna-based uncertainties to obtain these. These expressions make sense intuitively. If we ignore the cross-terms initially, the solution scales inversely with the total calibrator signal strength ($\sum B_i$). The amplitude precision depends solely on the antennas forming the baseline, whereas the phase precision depends on the relative contributions from all baselines involving the antennas in question. This is due to the phase being a relative quantity. Increasing the number of antennas improves the estimation precision, as does increasing the signal strengths. The cross-terms weight the contributions from individual antennas, according to the baseline projections on the vector between the calibrator sources.

Up to this point, the noise parameter, $\sigma$, has referred to the thermal noise, which, for the MWA 32T is $\sim$15 Jy per visibility in each coarse 1.28 MHz channel. However, the MWA will be confusion-limited, and the actual `system noise' will be higher due to the rms fluctuations generated by the confusing sources. Assuming a confusion of 1 Jy/beam, this corresponds to $\sim$100 Jy rms in each visibility. These `noise' terms are independent, and can be added in quadrature to produce an overall system noise of $\sim$100 Jy in each visibility.

\begin{figure}
\subfigure{\includegraphics[scale=0.4]{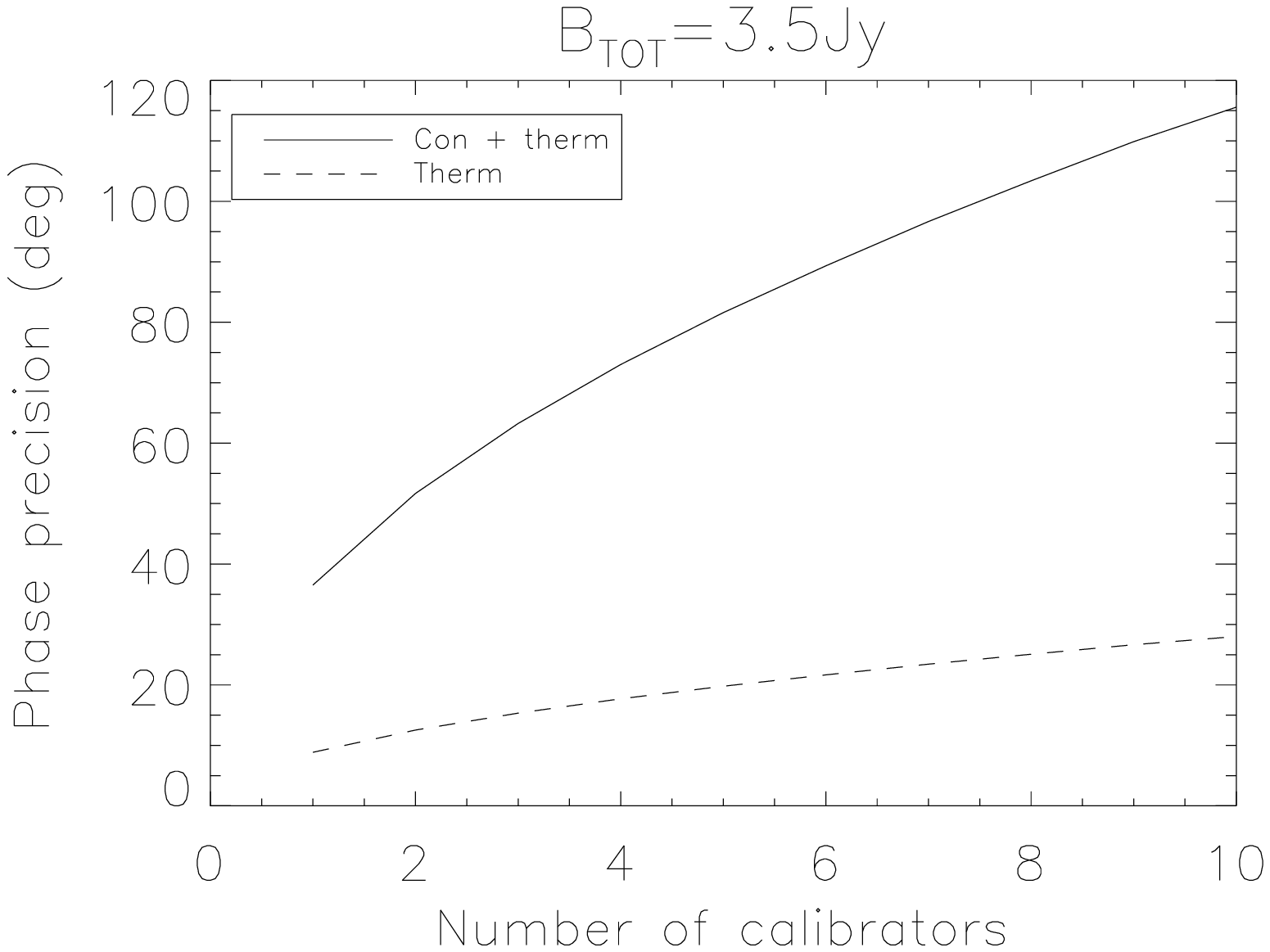}}
\subfigure{\includegraphics[scale=0.4]{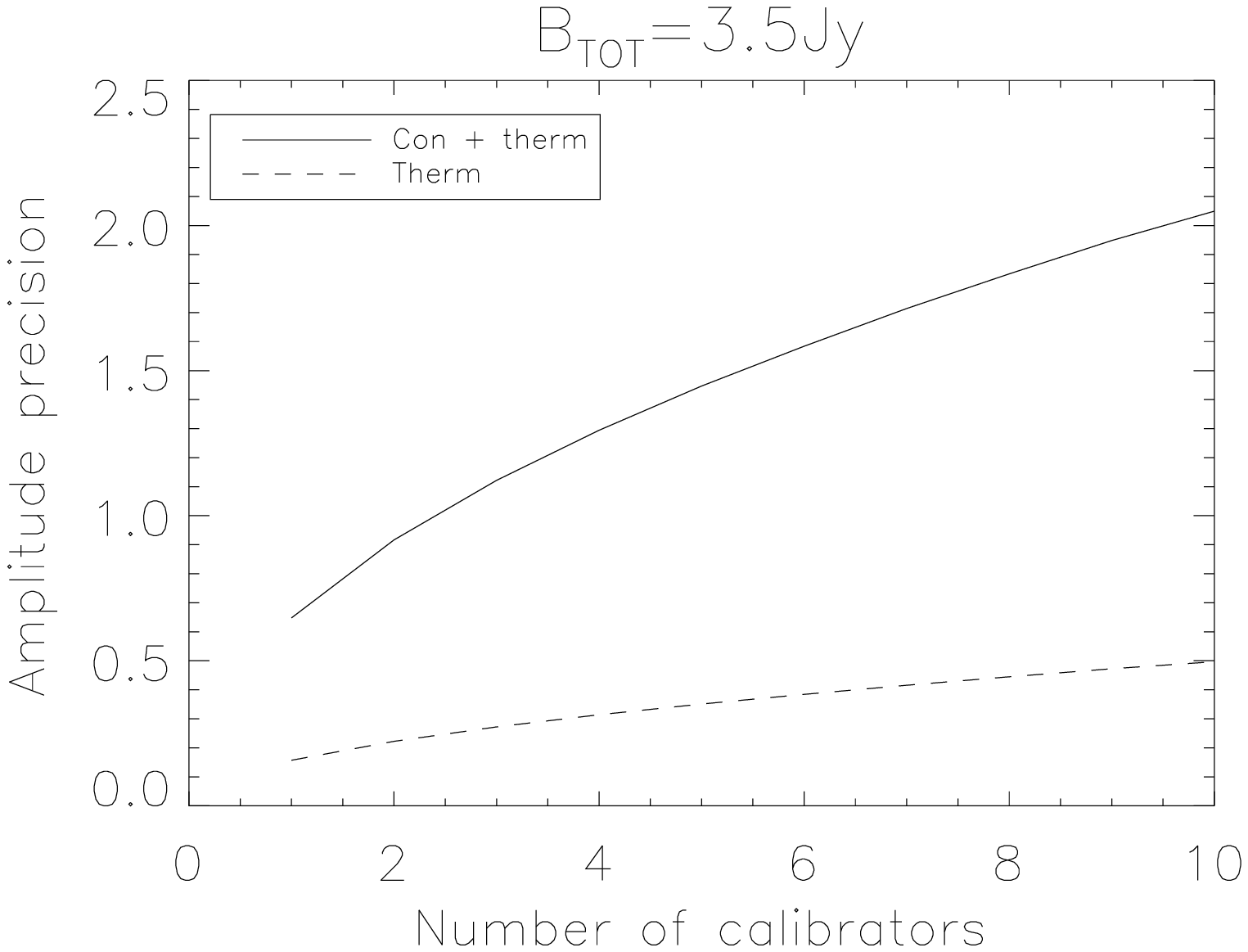}}\\
\subfigure{\includegraphics[scale=0.4]{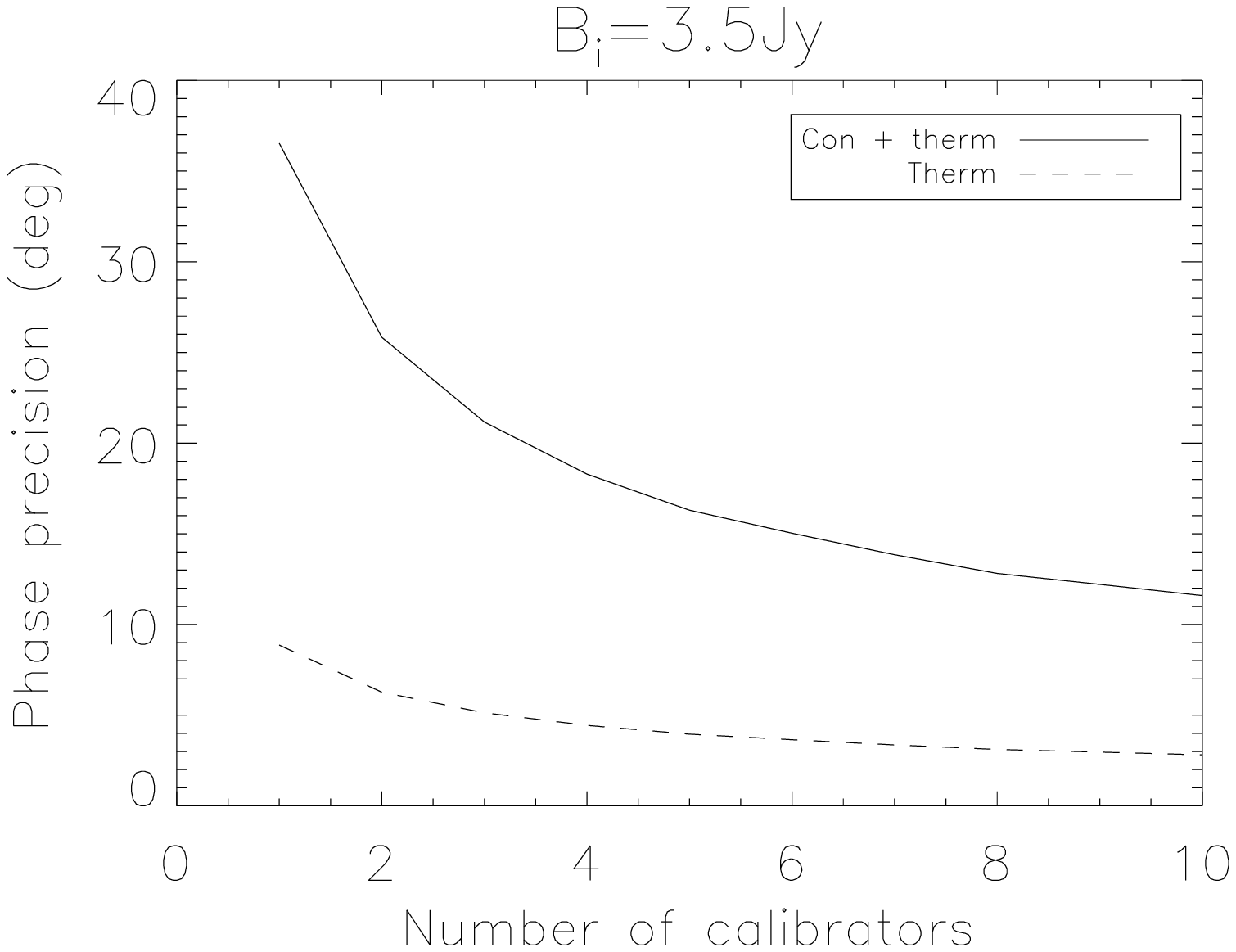}}
\subfigure{\includegraphics[scale=0.4]{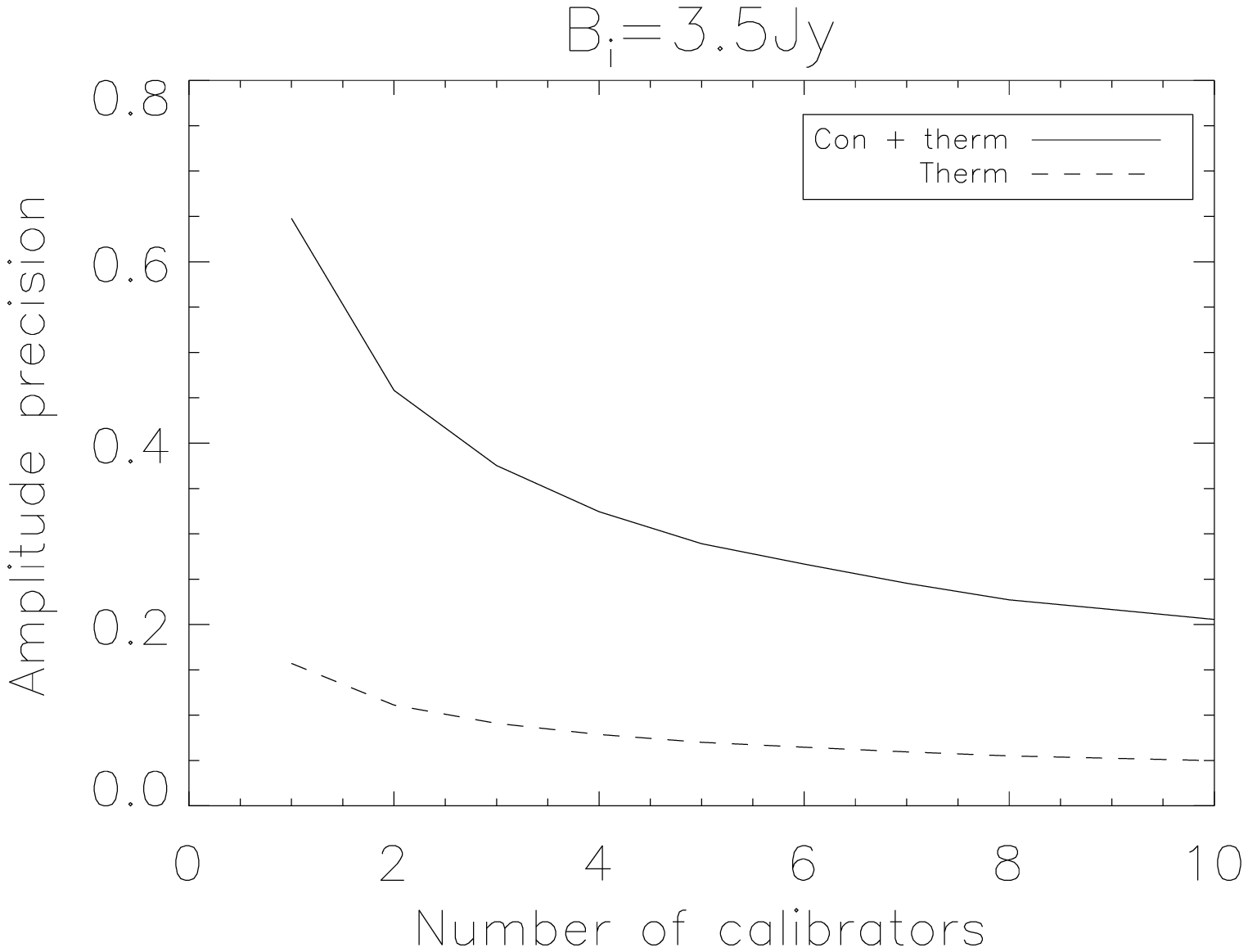}}
\vspace{1cm}
\caption{(Upper) Gain phase and amplitude precision for a single baseline, and all frequency channels, for a total calibrator flux density of 3.5Jy, as a function of number of calibrators ($\sigma=15$ Jy/channel/baseline, $\alpha=0$, $\Delta{t}=8$s, $F=24$, $\Delta{\nu}=30.72$ MHz). (Lower) Bounds as a function of number of calibrators, where each has a flux density of 3.5 Jy.}
\label{calibratorfig}
\end{figure}

In Figure \ref{calibratorfig} we show the calibration amplitude and phase estimation precision, for the current MWA 32-tile system, and for varying calibrator numbers and strengths. For each plot, the precision is shown for thermal noise alone (`Therm') and for thermal$+$confusion (`Con$+$therm').
The upper figures show the precision on phase and amplitude estimation for a single baseline, for a \textit{total} calibrator flux density of 3.5 Jy, and as a function of number of calibrators. These are ensemble averages to remove the effects of calibrator position. These figures demonstrate that it is advantageous to have a single bright calibrator, rather than a few lower signal-to-noise calibrators. The precision scales as the square-root of the number of calibrators. The lower figures show the precision as a function of number of calibrators, where each calibrator has $B$=3.5 Jy. This corresponds to S/N=5 for an 8s integration. The precision here scales as the inverse square-root of the number of calibrators. In all cases, the amplitude gain parameters ($b$) are set to unity. Note that these plots are for a particular baseline. In general, the plots could vary substantially between baselines. If, for example, there was little calibrator information for a particular antenna, the precision will be degraded for the visibilities across all of the baselines it forms. Therefore, the connectivity (correlation) of the antennas is naturally accounted for in this formalism.

\subsubsection{Inclusion of calibration uncertainty in estimation of source parameters}
Now we have quantified the uncertainty that field-based calibration introduces into the data, the next step is to incorporate the additional uncertainty in measuring the parameters of a source. Previously, the Cramer-Rao bounds on estimating source parameters assumed only thermal noise in the system. To this noise we now add components that quantify the additional uncertainty. There are two sets of uncertainties to be introduced: a covariance matrix, $\boldsymbol{C}_c$ that quantifies the amplitude uncertainty, and a set of phase parameters that quantify the phase uncertainty (broadening the overall likelihood function acts on the real and imaginary components of the data, and therefore cannot easily include errors on the phase).

For a general covariance matrix and complex data, the general expression for the Fisher Information Matrix becomes:
\begin{eqnarray}
[\boldsymbol{I(\theta)}]_{ij} = {\rm{tr}}\left[ \boldsymbol{C}^{-1}(\boldsymbol{\theta})\frac{\partial\boldsymbol{C}(\boldsymbol{\theta})}{\partial{\theta_i}}\boldsymbol{C}^{-1}(\boldsymbol{\theta})\frac{\partial\boldsymbol{C}(\boldsymbol{\theta})}{\partial{\theta_j}} \right] \\\nonumber
+ 2{\rm{Re}}\left[\frac{\partial{\tilde{\boldsymbol{s}}^H({\boldsymbol{\theta}})}}{\partial{\theta_i}}{\boldsymbol{C}^{-1}}(\boldsymbol{\theta})\frac{\partial{\tilde{\boldsymbol{s}}({\boldsymbol{\theta}})}}{\partial{\theta_j}}\right],
\end{eqnarray}
where we write the covariance matrix, $\boldsymbol{C}$, as:
\begin{equation}
\boldsymbol{C} = (\sigma^2 \boldsymbol{I} + \boldsymbol{C}_c)
\end{equation}
and $\boldsymbol{C}_c$ is the covariance matrix due to the amplitude calibration uncertainty. Note that the noise term, $\sigma$, is the thermal noise --- we assume that the background sources have been modelled and subtracted (including confusing sources). In practise, some level of background source will remain, and this also can be included in the modelling.

The construction of the covariance matrix, $\boldsymbol{C}_c$ reflects the calibration uncertainties on each baseline (equations \ref{b_precision}--\ref{calibration_precision}). For the pedagogical case we are considering here, we assume that the calibration process does not introduce any correlations between baselines or channels, other than through the common antenna gain term. This is equivalent to asserting that there are no baseline-based errors. In reality, the off-diagonal terms will be non-zero, but small. We write the $n$th component of the diagonal covariance matrix as:
\begin{equation}
[\boldsymbol{C}_c]_n = B^2\Delta{b}_n^2,
\end{equation}
where $B$ is the strength of the source and $\Delta{b}_n$ is the calibration amplitude precision for baseline $n$. The source strength is here to have the correct units, and reflects the dependence of the absolute scale of the calibration errors on the source strength. Using this scheme, the system noise for low signal-to-noise ratio sources will be dominated by the thermal noise, whereas high signal-to-noise sources will have a relatively larger calibration error component.

The phase uncertainty is modelled as an additional parameter, $\psi_n$, in the argument of the exponential in the signal. This is a random (as compared with a deterministic) parameter, for which we possess prior knowledge (the phase calibration uncertainty), and is dependent on baseline, $n$. Note that we are referring here to uncertainty in a statistical sense: we do not wish to estimate the actual phase fluctuations for each antenna (which are constrained by closure phase, and are nuisance parameters), but instead want to understand the additional uncertainty they introduce. Instead of estimating the four deterministic parameters of the transient source ($B,\alpha,l,m$), we simultaneously estimate these parameters and the $N$ random phase parameters, $\psi_n$. We use the prior knowledge of how these parameters are distributed to include additional information in a modified Fisher Information Matrix.

The CRLB cannot be extended easily to include prior information. An equivalent expression for random parameters is available using a Bayesian approach where the probability distribution function describing the data includes the probability distribution function of the parameter. This approach allows prior information on the value of the parameter to be incorporated into the bound. \citet{rockah87} introduced the Hybrid Cramer-Rao lower bound (HCRLB) as an extension to the CRLB that allows estimation of both random and deterministic parameters. The probability distribution functions of the random parameters can contain prior information on the distribution of that parameter, and improve the estimation performance. In practice, this is achieved by the Fisher Information containing contributions from both the data (classical CRLB) and the prior information.

The modified FIM is given by:
\begin{equation}
\boldsymbol{I}(\boldsymbol{\theta})^\prime = E_{\boldsymbol{\psi}}[\boldsymbol{I}(\boldsymbol{\theta})] + \boldsymbol{I}_{\rm pr}(\boldsymbol{\psi}),
\end{equation}
where $\boldsymbol{I}(\boldsymbol{\theta})$ is the classical (data) FIM, and $\boldsymbol{I}_{\rm pr}(\boldsymbol{\psi})$ is the prior information, and is given by:
\begin{equation}
E_{\boldsymbol{\psi}}\left[ \frac{\partial\log p(\boldsymbol{\psi})}{\partial \psi_i}^H\frac{\partial\log p(\boldsymbol{\psi})}{\partial \psi_j} \right]
\end{equation}
for component $ij$. The expectation over the random parameter of the data component is often omitted for tight prior distributions, resulting in a modified FIM that is the sum of the data and prior components. For gaussian PDFs with variance, $\sigma^2$, the prior information is:
\begin{equation}
\boldsymbol{I}_{\rm pr}(\psi_n) = 1/\sigma^2_{\psi_n}.
\end{equation}
As discussed above, there is no covariance between the source position parameters ($l,m$) and the source amplitude parameters ($B,\alpha$). The random phase parameters, $\psi_n$, also do not co-vary with the amplitude parameters, and their inclusion therefore has no impact on the ability to estimate them (the additional \textit{amplitude} uncertainty does affect all parameters, however). In the case of calibration phase uncertainty \textit{and} atmospheric phase noise, the prior PDF is broadened to include contributions from both uncertainties.

\begin{figure}
\subfigure[]{\includegraphics[scale=0.4]{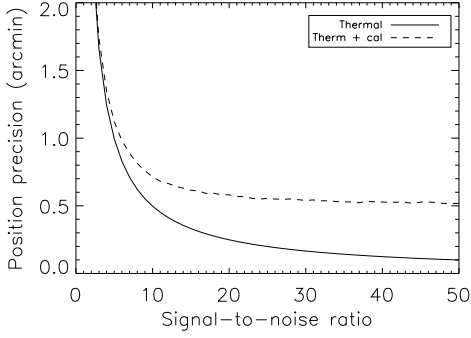}}
\subfigure[]{\includegraphics[scale=0.4]{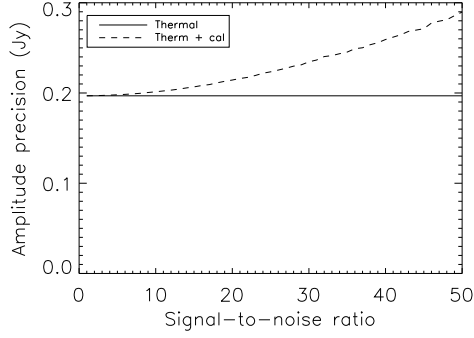}}\\
\subfigure[]{\includegraphics[scale=0.4]{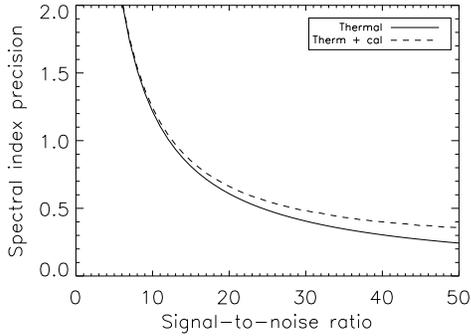}}
\subfigure[]{\includegraphics[scale=0.4]{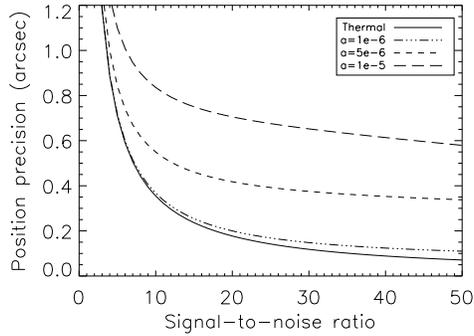}}
\vspace{1cm}
\caption{(a) Precision in estimating the source sky position ($l$) for a source as a function of signal-to-noise ratio, for calibration based on five 1 Jy (S/N=7) sources ($\sigma=15$ Jy/channel/baseline, $\alpha=0$, $\Delta{t}=8$s, $F=24$, $\Delta{\nu}=30.72$ MHz). (b) Amplitude ($B$) precision. (c) Spectral index ($\alpha$) precision. (d) Sky position ($l$) precision for ASKAP, including three magnitudes of atmospheric phase noise ($a$=10$^{-6}$, 5$\times$10$^{-6}$, 10$^{-5}$).}
\label{estimation_calibration}
\end{figure}

We now form the $(N+4) \times (N+4)$ FIM for the information carried in the data about the source parameters, and invert to yield the lower bounds on parameter estimates. This FIM now includes the effects of amplitude and phase calibration. Figures \ref{estimation_calibration}(a-c) display the maximum estimation precision for a source, as a function of source signal-to-noise ratio (thermal noise), for system noise being purely thermal, and for thermal$+$calibration errors. In these figures, the calibration is performed using five 1 Jy (each with S/N=7) sources in the field, and is performed every 8 seconds.
It is clear from these plots that the MWA 32T is dominated by thermal noise and confusion for this case, and that the field-based calibration is not a significant source of uncertainty for low signal-to-noise ratio sources. Of course, a reduction in the number or strength of calibrator sources will affect these results. As mentioned earlier, the number of calibrators considered here are those within some scale on the sky over which the atmosphere can be considered stable. This group of calibrators can then be used jointly to estimate the gains. For stable atmospheres, and large patches of sky, the number of calibrators will be large, and the calibration precision will be high. For an unstable atmosphere with small patches, the number of calibrators will be low, and the calibration precision will be low. Therefore, this formalism includes the effects of position-dependent calibration in a rudimentary way.

Figure \ref{estimation_calibration}(d) shows the ideal precision for estimation of source position ($l$) for ASKAP at 1.4GHz ($\Delta\nu$=300MHz, 32 channels, 5 second integration), and including three levels of atmospheric phase noise. In the case of the ASKAP system, the calibration uncertainty is small, and the additional uncertainty on the source position is dominated by the atmospheric phase noise.


\section{Optimal detector}\label{final_likelihood}
We have presented an analytical model for the impact of calibration, background confusing sources and the atmosphere on source estimation and precision. With this model, and the statistical framework developed in section \ref{sdt}, we can describe the optimal detector for visibility data. The signal-present hypothesis likelihood function for one integration timestep can be described by:
\begin{equation}
L(\tilde{\bf{x}};H_1) = \frac{1}{\pi^{FN}\det(\boldsymbol{C}_c+\sigma^{2}\boldsymbol{I})}
\exp{\left[-(\tilde{\bf{x}}-\tilde{\bf{s}})^H(\boldsymbol{C}_c+\sigma^{2}\boldsymbol{I})^{-1}(\tilde{\bf{x}}-\tilde{\bf{s}})\right]},
\label{cal_like}
\end{equation}
where $\boldsymbol{C}_c$ contains the amplitude calibration uncertainties, and the signal for channel $f$ and baselines $n$ is given by:

\tiny
\begin{equation}
\tilde{s}[f,n] = B(l,m)\left(\frac{\nu(f)}{\nu_0}\right)^{\alpha}\exp{[-2\pi{i}(u_{fn}l+v_{fn}m + \psi_n)]}+\displaystyle\sum_{k=1}^{K} B_k\exp{[-2{\pi}i(u_{fn}l_k+v_{fn}m_k+\psi_n)]},
\end{equation}
\normalsize
and the phase parameters are gaussian distributed according to the calibration and atmospheric phase uncertainties. The likelihood function for the signal-absent hypothesis, $H_0$, has the same form as equation \ref{cal_like}, and the `signal' is given by:
\begin{equation}
\tilde{s}[f,n] = \displaystyle\sum_{k=1}^{K} B_k\exp{[-2{\pi}i(u_{fn}l_k+v_{fn}m_k+\psi_n)]},
\end{equation}
(i.e., background static sources). Note that both likelihood functions require the phase uncertainty terms to be removed.

To perform the detection, one needs to estimate or remove all of the unknown parameters. The unknown source parameters are estimated using maximum likelihood estimation (ideally), and the random phase parameters are integrated out, using their prior PDFs and the Bayesian approach described in section \ref{bayesian}. Algorithmically, the unknown deterministic parameters are estimated first, with the phase parameters set at their mean values ($\psi_n=0$) for simplicity (in practise, the phase errors will be small, and this simplification will have minimal impact on the detection performance). Then, the $N$ one-dimensional integrals are performed to remove the $N$ random parameters. Finally, the GLRT is performed by taking the ratio of the values of the likelihood functions, and the result compared with a threshold. In Paper II we use the results derived here to form realistic likelihood functions, and present algorithms for implementing the optimal detector.

\section{Conclusions}
We have presented a framework for designing optimal source detectors with visibility-space data from interferometric arrays, and applied this to describe a realistic optimal detector. Working in visibility space allows a more natural characterisation of the data likelihood functions than image space, where noise is structured and not well-understood. Source detection is complicated by unknown source strength, spectral index, position, arrival time and duration (transient sources), and the uncertainty on these parameters reduces detection performance. Estimation of these parameters is required before signal detection can be performed. Uncertainty introduced by field-based calibration, confusing sources and atmospheric phase noise further complicates signal detection and reduces detection performance. We have explored the impact of these additional sources of uncertainty on the ability of an efficient estimator to determine the parameters of a source, and applied these methods to two SKA pathfinder instruments: the MWA 32T and ASKAP. We then used an understanding of these effects to present a realistic model of visibility data, and design an optimal detector.

\section*{Appendix}
\appendix
\section{Derivation of calibration precision}
\label{appendix_crb}
We have calculated (and will present below) the estimation precision for a single calibrator, and two calibrators. From this, we can extrapolate to $N_c$ calibrators, based on the form. The precision with which the calibration solution for a given \textit{baseline} can be theoretically measured (since this is the data we measure) can be calculated using error propagation from the bounds on the individual antennas alone (and the covariances). For a single calibrator, and three antennas, the precision with which the gain amplitude can theoretically be measured is given by:
\begin{equation}
\Delta{b}_{\alpha\beta} \geq \frac{\sigma}{B\sqrt{2\displaystyle\sum_{f=1}^F \left(\frac{\nu(f)}{\nu_0}\right)^{2\alpha}}},
\end{equation}
i.e., the inverse signal-to-noise for complex data. For two calibrators, the expression becomes:
\begin{eqnarray}
\Delta{b}_{\alpha\beta} &\geq& \frac{\sigma}{\sqrt{2}} \left[ \displaystyle\sum_{f=1}^F\left(B_1^2\left(\frac{\nu(f)}{\nu_0}\right)^{2\alpha_1}+B_2^2\left(\frac{\nu(f)}{\nu_0}\right)^{2\alpha_2} 
 \right. \right. \nonumber \\  &\null&  \qquad \left. \left. 
+2B_1B_2\left(\frac{\nu(f)}{\nu_0}\right)^{\alpha_1+\alpha_2}\cos{2\pi(u_{f\alpha\beta}(l_2-l_1)+v_{f\alpha\beta}(m_2-m_1))}\right) \right]^{-1/2}.
\end{eqnarray}
Extrapolating to $N_c$ calibrators yields:
\begin{eqnarray}
\Delta{b}_{\alpha\beta} &\geq& \frac{\sigma}{\sqrt{2}} \left[ \displaystyle\sum_{f=1}^F\left(\displaystyle\sum_{i=1}^{N_c} B_i^2\left(\frac{\nu(f)}{\nu_0}\right)^{2\alpha_i} 
\right. \right. \nonumber \\  &\null&  \qquad \left. \left. 
+ \displaystyle\sum_{i=1}^{N_c}\displaystyle\sum_{j\neq{i}}^{N_c}B_iB_j\left(\frac{\nu(f)}{\nu_0}\right)^{\alpha_i+\alpha_j} \cos{2\pi(u_{f\alpha\beta}(l_j-l_i)+v_{f\alpha\beta}(m_j-m_i))}\right) \right]^{-1/2},
\end{eqnarray}
which, for $N_c$ identical, co-located calibrators with strength $B$ and $\alpha=0$, gives $\sigma/(\sqrt{2F}N_cB)$.

For the gain phase precision (setting $\phi_1$ to zero) and one calibrator, the precision is:
\begin{equation}
\Delta\phi_{\beta\gamma} \geq \frac{\sigma}{\sqrt{8\pi^2\displaystyle\sum_{f=1}^F \left(\frac{\nu(f)}{\nu_0}\right)^{2\alpha}}}\frac{\sqrt{b_\beta^2+b_\gamma^2}}{Bb_\beta{b}_\gamma}\frac{1}{\sqrt{b_1^2+b_2^2+b_3^2}}.
\end{equation}
For two calibrators, this expands to include the cosine cross-terms, and is given by:
\begin{equation}
\Delta\phi_{12} \geq \frac{\sigma}{\sqrt{8\pi^2}}
\frac{\sqrt{b_1^2X_{13}+b_2^2X_{23}}}{b_1b_2}\frac{1}{\sqrt{b_1^2X_{12}X_{13}+b_2^2X_{12}X_{23}+b_3^2X_{13}X_{23}}}.
\end{equation}
where

\tiny
\begin{equation}
X_{ab} = \displaystyle\sum_{f=1}^F \left[B_1^2\left(\frac{\nu(f)}{\nu_0}\right)^{2\alpha_1}+B_2^2\left(\frac{\nu(f)}{\nu_0}\right)^{2\alpha_2}+2B_1B_2\left(\frac{\nu(f)}{\nu_0}\right)^{\alpha_1+\alpha_2}\cos{2\pi(u_{ab}(l_2-l_1)+v_{ab}(m_2-m_1))}\right].
\end{equation}
\normalsize
Extrapolating to $N_c$ calibrators gives:
\begin{equation}
\Delta\phi_{12} \geq \frac{\sigma}{\sqrt{2(2\pi)^{2}}}
\frac{\sqrt{b_1^2X_{13}+b_2^2X_{23}}}{b_1b_2}\frac{1}{\sqrt{b_1^2X_{12}X_{13}+b_2^2X_{12}X_{23}+b_3^2X_{13}X_{23}}}.
\end{equation}
where
\begin{equation}
X_{ab} = \displaystyle\sum_{f=1}^F \left[\displaystyle\sum_{i=1}^{N_c} B_i^2\left(\frac{\nu(f)}{\nu_0}\right)^{2\alpha_i} + \displaystyle\sum_{i=1}^{N_c}\displaystyle\sum_{j\neq{i}}^{N_c}B_iB_j\left(\frac{\nu(f)}{\nu_0}\right)^{\alpha_i+\alpha_j} \cos{2\pi(u_{ab}(l_j-l_i)+v_{ab}(m_j-m_i))}\right].
\label{Xeqn}
\end{equation}
Note that this is still the solution for a three-antenna system. The general solutions for $N_c$ calibrators and $M$ antennas are:
\begin{eqnarray}
\Delta{b}_{\beta\gamma} &\geq& \frac{\sigma}{\sqrt{2}} \left[ \displaystyle\sum_{f=1}^F\left(\displaystyle\sum_{i=1}^{N_c} B_i^2\left(\frac{\nu(f)}{\nu_0}\right)^{2\alpha_i} 
\right. \right. \nonumber \\  &\null&  \qquad \qquad \left. \left. 
+ \displaystyle\sum_{i=1}^{N_c}\displaystyle\sum_{j\neq{i}}^{N_c}B_iB_j\left(\frac{\nu(f)}{\nu_0}\right)^{\alpha_i+\alpha_j} \cos{2\pi(u_{f\beta\gamma}(l_j-l_i)+v_{f\beta\gamma}(m_j-m_i))}\right) \right]^{-1/2} \\
%
\Delta\phi_{\alpha\beta} &\geq& \frac{\sigma}{\sqrt{2(2\pi)^{2}}}
\sqrt{\varepsilon_{ab...z\neq\alpha\neq\beta}A_{1,a}A_{2,b}...A_{M-1,z}}
\Bigg{/}
\sqrt{\varepsilon_{ab...z}A_{1,a}A_{2,b}...A_{M-1,z}},
\label{calibration_precision_app}
\end{eqnarray}
where $\varepsilon$ is the Levi-Civita permutation symbol, $(a,b,...,z)\in{[1,M-1]}$, and $A_{a,b}$ are the FIM matrix elements, and are given by,
\begin{equation}
A_{a,b}=\left\{\begin{array}{cl}
	b_a^2\displaystyle\sum_{k\neq{a}}^{M-1}b_k^2X_{ak}, & a=b \\
	-b_a^2b_b^2X_{ab}, & a\neq{b}
	   \end{array}\right.,
\end{equation}
where $X_{ab}$ is the same as in equation \ref{Xeqn}. There is an implicit summation over all indices in equation \ref{calibration_precision_app}, \textit{viz},
\begin{equation}
\varepsilon_{ab...z}A_{1,a}A_{2,b}...A_{M-1,z} = \displaystyle\sum_{a=1}^{M-1}\displaystyle\sum_{b=1}^{M-1}...\displaystyle\sum_{z=1}^{M-1} \varepsilon_{ab...z}A_{1,a}A_{2,b}...A_{M-1,z}.
\end{equation}
The expression for the phase uncertainty is not easy to implement, and in practise, it is far simpler to form the FIM and invert numerically, and extract the baseline-based uncertainties using error propagation on the inverse FIM elements.

\section*{Acknowledgments} 
We would like to thank Matthew Whiting for providing the ASKAP antenna specifications and system characteristics. We would also like to thank the anonymous referee for providing a very considered and constructive review of the manuscript. Their input has improved the manuscript considerably.

\bibliographystyle{jphysicsB}
\bibliography{paper.bib}

\end{document}